\documentclass[aps,prd,nofootinbib,notitlepage,tightenlines,nofootinbib,twocolumn,superscriptaddress,longbibliography]{revtex4-2}
\usepackage{amsmath}
\usepackage{amssymb,amsthm}
\usepackage{mathtools}
\usepackage{mathrsfs}
\usepackage{relsize}
\usepackage{bm}
\usepackage{ulem}
\usepackage{epsfig,graphics,graphicx,color,xcolor}
\usepackage{soul} 
\usepackage{cancel} 
\usepackage{slashed}	
\usepackage{subfigure}
\usepackage{array}
\usepackage{ragged2e}
\usepackage{lineno}
\usepackage{natbib}
\usepackage{hyperref}
\hypersetup{colorlinks=true,linkcolor=red,anchorcolor=red,citecolor=orange, filecolor=brown,urlcolor=red,bookmarksnumbered=true,
pdfview=FitB
}
\graphicspath{ {image/} }
\usepackage{multirow}
\usepackage{enumitem}

\colorlet{darkgreen}{green!50!black}
\colorlet{darkblue}{blue!70!black}
\colorlet{brightyellow}{yellow!75!red}
\colorlet{orange}{red!50!yellow}
\colorlet{darkgray}{gray!50!black}

\def\fauxschelper#1 #2\relax{%
  \fauxschelphelp#1\relax\relax%
  \if\relax#2\relax\else\ \fauxschelper#2\relax\fi%
}
\def\Hscale{.85}\def\Vscale{.74}\def\Cscale{1.12}
\def\fauxschelphelp#1#2\relax{%
  \ifnum`#1>``\ifnum`#1<`\{\scalebox{\Hscale}[\Vscale]{\uppercase{#1}}\else%
    \scalebox{\Cscale}[1]{#1}\fi\else\scalebox{\Cscale}[1]{#1}\fi%
  \ifx\relax#2\relax\else\fauxschelphelp#2\relax\fi}
  
\def\dd{{\mathrm{d}}}

\newcommand{\half}[1][1] {\mathsmaller{\frac{#1}{2}}}

\makeatletter
\newcommand*{\transpose}{%
  {\mathpalette\@transpose{}}%
}
\newcommand*{\@transpose}[2]{%
  \raisebox{\depth}{$\m@th#1\intercal$}%
}
\makeatother

\begin{document}

\title{Two-photon transitions of charmonia on the light front}

\author{Yang~Li}
\affiliation{Department of Modern Physics, University of Science and Technology of China, Hefei 230026, China}

\author{Meijian Li}
 \email{mliy@jyu.fi}
\affiliation{Department of Physics, P.O. Box 35, FI-40014 University of Jyväskylä,
Finland}
\affiliation{
Helsinki Institute of Physics, P.O. Box 64, FI-00014 University of Helsinki,
Finland
}
\affiliation{
Instituto Galego de F\'\i sica de Altas Enerx\'\i as (IGFAE), Universidade de Santiago de Compostela, E-15782 Galicia, Spain
}

\author{James~P.~Vary}
\affiliation{Department of Physics and Astronomy, Iowa State University, Ames, IA 50011}

\date{\today}

\begin{abstract}
We investigate the two-photon transitions $H_{c\bar c} \to \gamma^*\gamma$ of the charmonium system in light-front dynamics. The light-front wave functions were obtained from solving the effective Hamiltonian based on light-front holography and one-gluon exchange interaction within the
basis light-front quantization approach. We compute the two-photon transition form factors as well as the two-photon decay widths for S- and P-wave charmonia, $\eta_c$ and $\chi_{cJ}$ and their excitations. Without introducing any free parameters, our predictions are in good agreement with the recent experimental measurements by BaBar and Belle, shedding light on the relativistic nature of charmonium. 
\end{abstract}
\maketitle 

\section{Introduction}

Charmonium is an intriguing system with entangled scales $\Lambda_\textsc{qcd} \lesssim \alpha_s m_c c^2\ll m_c c^2$ \cite{Brambilla:2010cs}. Typical estimates put the average velocity of the quarks $v^2_c\sim 0.3$. Thus, there may be large relativistic corrections for observables sensitive to short-distance physics. The two-photon transition of charmonium, viz. $H_{c\bar c} \to \gamma\gamma$, is one of the leading examples (see Refs.~\cite{Berger:1986ii, Poppe:1986dq, Feindt:1991rb, Lansberg:2019adr} for reviews), where predictions based on non-relativistic dynamics appear deficient. Typical symptoms include the slow convergence in non-relativistic effective theories (e.g., NRQCD \cite{Feng:2015uha, Feng:2017hlu}),  large differences among various non-relativistic potential model calculations and discrepancies with the experimental measurements \cite{Babiarz:2019sfa, Babiarz:2020jkh}. Driven by recent progress on experimental measurements, the call for fully relativistic approaches is clear.

Lattice computation of the two-photon transition with off-shell photons is particularly challenging, as demonstrated by the large discrepancy between theoretical predictions and the PDG value, since the energetic virtual photon is not an eigenstate of the QCD Hamiltonian \cite{Liu:2020qfz}. Nevertheless, strides have been made to access the two-photon width \cite{Dudek:2006ut, CLQCD:2016ugl, CLQCD:2020njc, Meng:2021ecs,Zou:2021mgf}, and transition form factors \cite{Dudek:2006ut, CLQCD:2016ugl}. Calculations from other relativistic approaches, notably DSE/BSE, have also been reported in the literature and appear successful \cite{Chen:2016bpj}. 

In this work, we report the calculation of the two-photon transition form factors (TFFs) of charmonium in a light-front Hamiltonian approach. This approach solves for the light-front wave functions (LFWFs) directly from a low-energy relativistic effective Hamiltonian for QCD \cite{Brodsky:1997de}. The advantage of this approach is evident from the short-distance $z^2 \sim 1/Q^2 \ll \Lambda^{-2}_\textsc{qcd}$ behavior of the transition amplitude $\int \dd^4x e^{iq\cdot z}\langle 0 | T\{J^\mu(z) J^\nu(0)\} | H(p) \rangle$, where the leading contribution comes from the light-cone distribution amplitude (LCDA) $\phi_P$ \cite{Lepage:1980fj, Chernyak:1983ej},
\begin{equation} 
F_{P\gamma\gamma}(Q^2) \overset{Q^2\gg \Lambda^2_\textsc{qcd}}{=} \int_0^1 \dd x \, T_H(x, Q) \phi_P(x, \tilde Q).
\end{equation}
The hard kernel $T_H = e_f^2f_P/(x(1-x)Q^2)$ is computable from perturbation theory. 

This result can be extended to the full range of $Q^2$ by using the LFWFs $\psi_P(x, \vec k_{\perp})$ \cite{Lepage:1980fj},
\begin{equation} 
F_{P\gamma\gamma}(Q^2) = \int_0^1 \dd x \int \frac{\dd^2k_\perp}{16\pi^3}\, T_H(x, \vec k_\perp, Q) \psi_P(x, \vec k_\perp),
\end{equation}
where the hard kernel $T_H$ is known to the next-to-leading order \cite{Beuf:2016wdz}. {The LFWFs and the associated LCDAs play a dominant role in exclusive processes \cite{Lepage:1980fj, Chernyak:1983ej, Chernyak:2014wra}. } 
The LFWFs adopted in this work were previously obtained in a light-front Hamiltonian approach to the charmonium spectra \cite{Li:2017mlw, Li:2017data}. 

We compute the two-photon widths and TFFs of the pseudoscalar $\eta_c$, scalar $\chi_{c0}$, axial vector $\chi_{c1}$, and tensor $\chi_{c2}$ and compare with the recent experimental data wherever available \cite{BaBar:2010siw, Belle:2017xsz, Belle:2020ndp, ParticleDataGroup:2020ssz}. The same LFWFs have been used to compute 
the decay constants, radiative transitions, semi-leptonic transitions, form factors, (generalized) parton distributions and cross sections of diffractive vector meson production with reasonable agreement over a remarkable wide range of experimental measurements and theoretical predictions \cite{Li:2017mlw, Li:2018uif, Tang:2020org, Adhikari:2018umb, Lan:2019img, Chen:2016dlk, Chen:2018vdw}. All these results, as well as those of the present work, represent predictions of the original model Hamiltonian for charmonia without adjusting its parameters \cite{Li:2017mlw}. 

\begin{figure}
\centering
\includegraphics[width=0.45\textwidth]{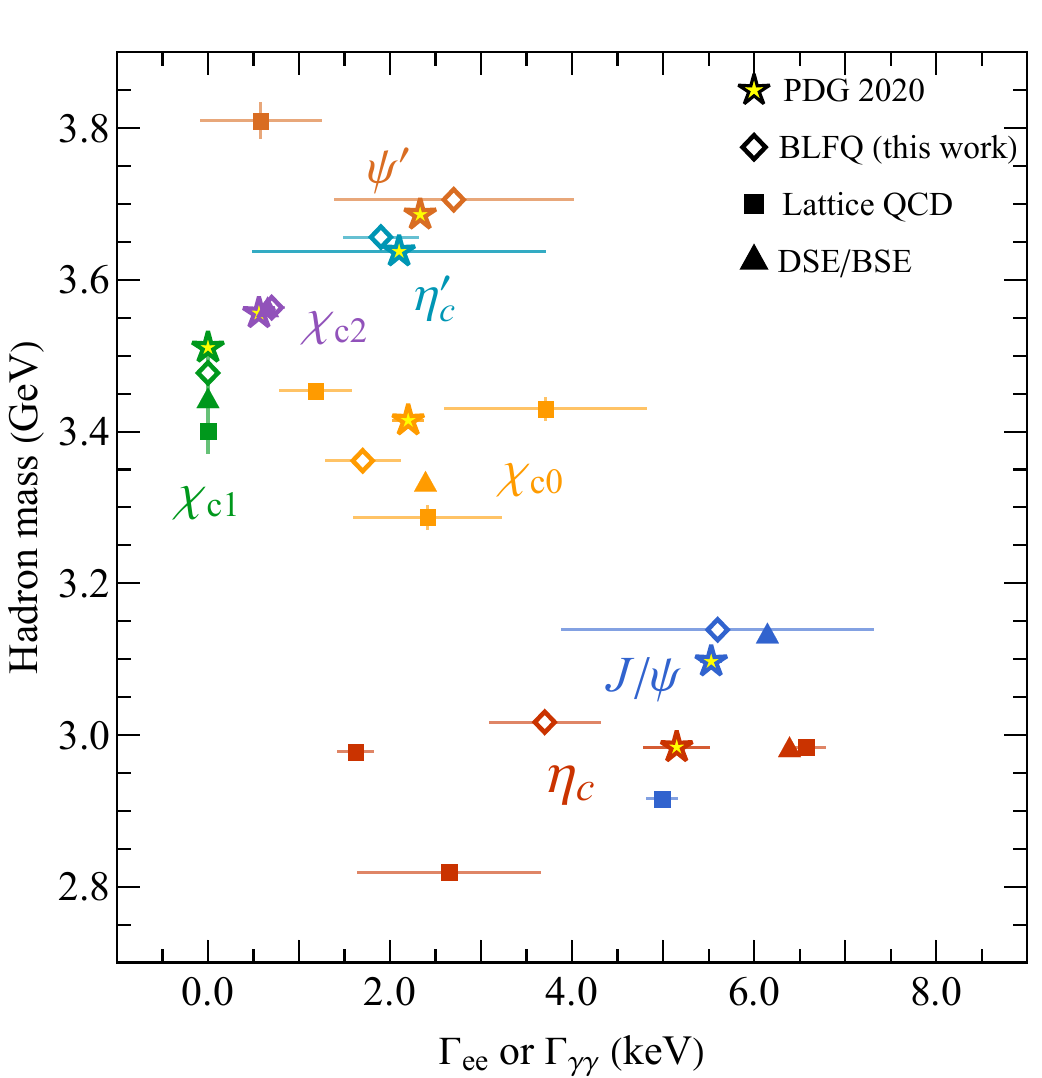}
\caption{(Colors online) The BLFQ prediction of charmonia mass and dilepton or diphoton width as compared with the PDG values as well as with the predictions from various Lattice QCD and DSE/BSE. }
\label{fig:combined}
\end{figure}

A comparative summary of our results for the combined predictions of the charmonium masses and dilepton (for vectors) or diphoton (for the rest) widths is shown in Fig.~\ref{fig:combined}. The plotted results are provided in Table~\ref{tab:TFF0}. 
The dilepton width $\Gamma_{ee}$ and the diphoton width $\Gamma_{\gamma\gamma}$ probe the similar physics since both quantities are proportional to the wave function at origin in the nonrelativistic quark model. Note that the diphoton width of $\chi_{c1}$ vanishes due to the Landau-Yang theorem \cite{Landau:1948kw, Yang:1950rg}.

\section{Formalism}

\begin{figure}
\centering
\includegraphics[width=0.5\textwidth]{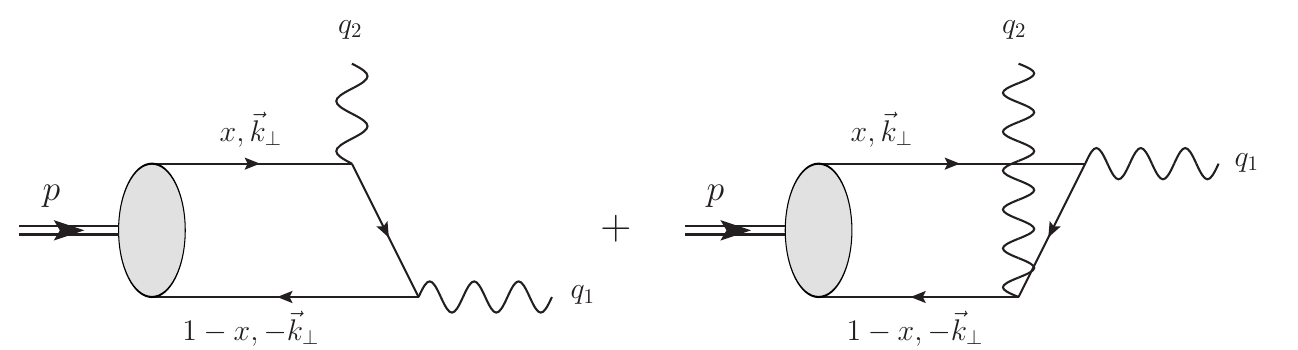}
\caption{Leading order diagrams of the transition form factor $\gamma^*\gamma \to H$.}
\label{fig:TFF_diagram}
\end{figure}

The leading-order contribution of the two-photon transition amplitude $\mathcal M^{\mu\nu\alpha}$ is shown in Fig.~\ref{fig:TFF_diagram}. 
It is convenient to introduce the associated helicity amplitudes, $H_{\lambda_1\lambda_2;\lambda} = \varepsilon^*_\mu(q_1,\lambda_1)\varepsilon^*_\nu(q_2,\lambda_2)e_\alpha(p,\lambda)\mathcal M^{\mu\nu\alpha}$ (for scalars and pseudoscalars, the hadron polarization tensor $e_\alpha = 1$), and evaluate the helicity amplitude 
in terms of the local hadronic matrix element, 
\begin{multline} \label{eq:def_helicity_amplitude}
H_{\lambda_1\lambda_2;\lambda} = \varepsilon^*_\nu(q_2,\lambda_2)\langle \gamma^*(q_1,\lambda_1)|J^\nu(0)|H(p,\lambda)\rangle.
\end{multline}
 
We choose a frame in which the light-cone dominance is manifest,
$q_1 = (q_1^+, 0, \vec q_{1\perp})$, $q_2 = (0, q_2^-, \vec q_{2\perp})$ \cite{Lepage:1980fj, Babiarz:2019sfa}. From momentum conservation,
$p = (q_1^+, q_2^-, \vec q_{1\perp}+\vec q_{2\perp})$. Here we adopt light-cone coordinate, 
$v = (v^+, v^-, v^1, v^2)$ where $v^\pm = v^0 \pm v^3$, and $\vec v_\perp = (v^1, v^2)$. 
The experimentally relevant case involves at least one photon on-shell. We choose the momentum of the on-shell photon to be $q_2$.
 
The helicity amplitudes can be expressed in terms of the LFWFs as (see Fig.~\ref{fig:TFF_diagram}),
\begin{align}\label{eq:LFWF_helicity_amplitude}
& H_{\lambda_1\lambda_2;\lambda}
= 
ee_f\sum_{s,\bar s} \int_0^1\frac{\dd x}{2x(1-x)}\int\frac{\dd^2k_\perp}{(2\pi)^3}  \nonumber \\
& 
\times \Big\{ \frac{1}{x} \sum_{s'}\psi_{s\bar s/\gamma}^{\lambda_1*}(x, \vec k_\perp)
\psi_{s'\bar s/H}^{\lambda}(x, \vec k'_\perp) 
\bar u_{s'}(k'_q) \slashed{\varepsilon}^*_2 u_{s}(k_q)  \nonumber\\
& - \frac{1}{1-x}  \sum_{\bar s'}\psi_{s\bar s/\gamma}^{\lambda_1*}(x, \vec k_\perp)
\psi_{s\bar s'/H}^{\lambda}(x, \vec k''_\perp) \bar v_{\bar s}(k_{\bar q}) \slashed{\varepsilon}^*_2 v_{\bar s'}(k'_{\bar q}) \Big\},
\end{align}
where, $\psi_{s\bar s/\gamma}$ and $\psi_{s\bar s/H}$ are the photon and meson LFWFs, respectively. The former can be computed using light-front perturbation theory \cite{Lepage:1980fj, Babiarz:2019sfa, Lappi:2020ufv}. 
The momenta of the quark  
before and after {the} photon emission are, $k_{q} = (xp^+, k^-_{q}, \vec k_\perp+x\vec p_{\perp})$, $k'_{q} = (xq_1^+, k'^-_{q}, \vec k'_\perp+x\vec q_{1\perp})$, 
respectively, where $\vec k'_\perp = \vec k_\perp+(1-x)\vec q_{2\perp}$. Note that only the light-front 3-momenta need to be specified since partons are always on their mass shells. 
Similarly, the momenta of the antiquark before and after {the} photon emission are,
$k_{\bar q} = \big((1-x)p^+, k^-_{\bar q}, -\vec k_\perp+(1-x)\vec p_{\perp}\big)$, $k'_{\bar q} = \big((1-x)p^+, k'^-_{\bar q},-\vec k''_\perp+(1-x)\vec q_{1\perp}\big)$, respectively, where $\vec k''_\perp = \vec k_\perp - x \vec q_{2\perp}$. 
 
In the single-tag case, only one photon is on-shell and the helicity amplitude $H_{\lambda_1\lambda_2;\lambda}$ can be extracted from the transverse current $\vec J_\perp$, similar to the M1 transition investigated in Ref.~\cite{Li:2018uif}. 

\subsection{$\eta_c \to \gamma\gamma$}

The two-photon transition amplitude of a pseudoscalar can be parametrized by a single form factor \cite{Hoferichter:2020lap}, 
\begin{equation}\label{eqn:TFFPg}
\mathcal M^{\mu\nu} =
4\pi\alpha_\mathrm{em} \varepsilon^{\mu\nu\rho\sigma} q_{1\rho} q_{2\sigma} F_{P\gamma\gamma}(q^2_1, q^2_2).
\end{equation}
For the important case where one of the photons is on-shell, it is useful to define a single-variable TFF $F_{P\gamma}(q^2) \equiv F_{P\gamma\gamma}(-q^2, 0) = F_{P\gamma\gamma}(0, -q^2)$.
$F_{P\gamma}(q^2)$ is related to the two-photon width, 
\begin{equation}\label{eqn:width_P}
\Gamma_{P\to\gamma\gamma} = \frac{\pi}{4} \alpha_\mathrm{em}^2 M_P^3 \big|F_{P\gamma}(0)\big|^2.
\end{equation}

\begin{figure}
\centering
\subfigure[\ $\eta_c(1S)$ \label{fig:TFF_etac_1S}]{\includegraphics[width=0.48\textwidth]{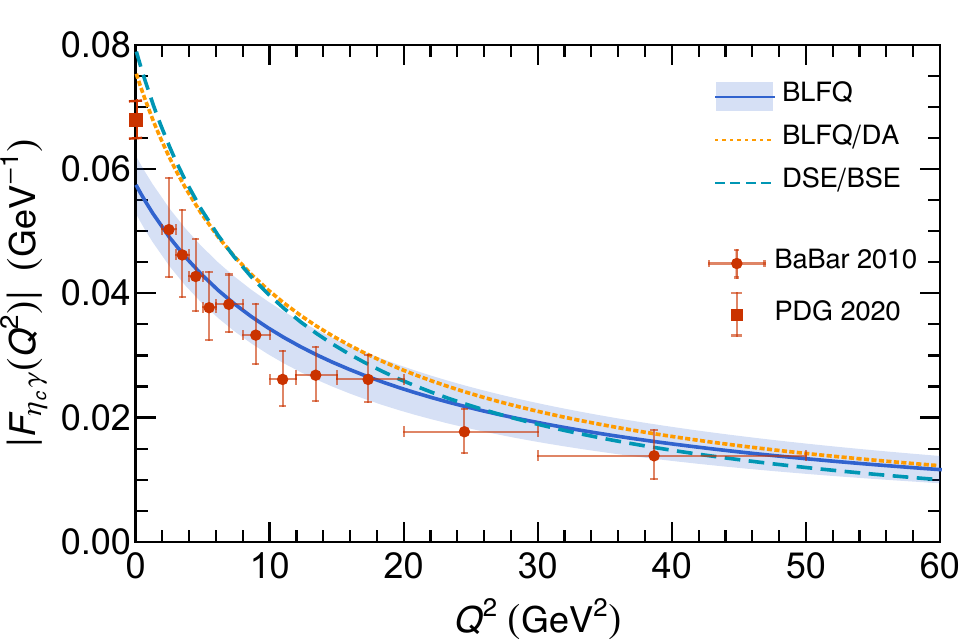}} \quad
\subfigure[\ $\eta_c(2S)$ \label{fig:TFF_etac_2S}]{\includegraphics[width=0.48\textwidth]{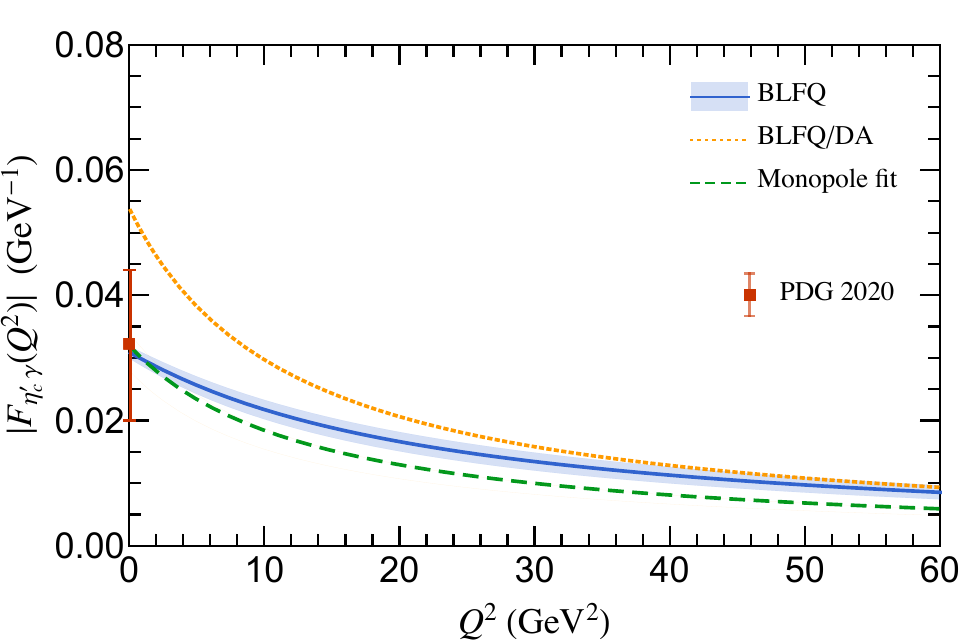}}
\caption{The two-photon transition form factor of charmed pseudoscalar mesons (a) $\eta_c(1S)$ and (b) $\eta_c(2S)$. The BLFQ predictions employs (\ref{eqn:LFWF_TFFPg}) whereas 
the BLFQ/DA prediction employs (\ref{eqn:DA_TFFPg}). For $\eta_c(1S)$ the result is compared with calculations from DSE/BSE \cite{Chen:2016bpj}. A monopole fit with pole mass $\Lambda = M_{\psi(2S)}$ is provided for $\eta_c(2S)$ as a reference (see texts). }
\label{fig:TFF_etac}
\end{figure}

The matrix element associated with (\eqref{eqn:TFFPg}) has the structure of pseudoscalar-vector-vector (PVV) coupling, where each vector meson is substituted by a virtual photon. Using the techniques developed in Ref.~\cite{Li:2018uif} for the M1 transition form factor $V_{P\to V\gamma}$, we obtain a LFWF representation  from the helicity amplitude $H_{0,\pm;0}$ (viz. matrix element of $J_\perp$) \cite{Babiarz:2019sfa}, 
\begin{multline}\label{eqn:LFWF_TFFPg}
F_{P\gamma}(Q^2) = e_f^2 2\sqrt{2N_C} \int \frac{\dd x}{2\sqrt{x(1-x)}} \int \frac{\dd^2k_\perp}{(2\pi)^3} \\
\times \frac{\psi_{\uparrow\downarrow-\downarrow\uparrow/P}(x, \vec k_\perp)}{k_\perp^2+m_f^2+x(1-x)Q^2}, 
\end{multline}
where $m_f$ is the quark mass, $N_C = 3$, and $e_f = 2/3$. 
In the above expression for the single-tag TFF, terms like $\vec q_{2\perp} \cdot \vec k_\perp$ vanish. 
Note that this decoupling is not generally held when both photons are off-shell.
Alternatively, one can also extract the TFF from the helicity amplitude $H_{\pm, 0; 0}$, i.e., matrix element of $J^+$. 
This choice leads to a slightly different expression, and the difference is expected to vanish in the NR limit~\cite{Li:2021cwv}. 
As a numerical example, the result differs from Eq.~\eqref{eqn:LFWF_TFFPg} at most 6\% over the range of $Q^2$ from 0 to 60 $\mathrm{GeV}^2$ for the $\eta_c$ TFF shown in Fig.~\ref{fig:TFF_etac_1S} .

At large $Q \gg \max\{\Lambda_\textsc{qcd}, m_f\}$, Eq.~(\ref{eqn:LFWF_TFFPg}) reduces to the celebrated partonic interpretation of Brodsky-Lepage \cite{Lepage:1980fj}, 
\begin{equation}\label{eqn:DA_TFFPg_BL}
F_{P\gamma}(Q^2) = \frac{e_f^2 f_P}{Q^2} \int_0^1 \dd x \frac{\phi_P(x, Q)}{x(1-x)}, 
\end{equation}
where, $\phi$ is the (normalized) pseudoscalar LCDA, and $f_P$ is the pseudoscalar decay constant. Their relations with the LFWFs are \cite{Li:2017mlw},
\begin{equation}\label{eqn:LCDA}
f_P \phi_P(x, \mu) =  
 \sqrt{\frac{2N_C}{x(1-x)}} \int\limits^{\mu^2} \frac{\dd^2k_\perp}{(2\pi)^3} {\psi_{\uparrow\downarrow-\downarrow\uparrow/P}(x, \vec k_\perp)},
\end{equation}
and 
$\int_0^1 \phi_P(x, \mu) =  1$.

For heavy quarkonia, since $m_f \gg \Lambda_\textsc{qcd}$, Eq.~(\ref{eqn:DA_TFFPg_BL}) can be extended to moderate $Q^2$ with $x(1-x)Q^2+m_f^2 \gg \langle k^2_\perp\rangle$ \cite{Babiarz:2019sfa} (cf. \cite{Hoferichter:2020lap}),
\begin{equation}\label{eqn:DA_TFFPg}
F_{P\gamma}(Q^2) = {e_f^2 f_P} \int_0^1 \dd x \frac{\phi_P(x, Q)}{x(1-x)Q^2 + m_f^2}. 
\end{equation}
 
This expression at $Q^2=0$ implies ($2m_f \approx M_P$), 
\begin{equation}\label{eqn:width_P_approx}
\Gamma_{P\to\gamma\gamma} \overset{.}{=} \frac{\pi e_f^4}{4}  \alpha_\mathrm{em}^2M_P^3\frac{f_P^2}{m_f^4} \approx 4 \pi e_f^4 \alpha_\mathrm{em}^2 \frac{f_P^2}{M_P}.
\end{equation}
However, this is only accurate in the non-relativistic limit. At small $Q^2$, these two equations~(\ref{eqn:DA_TFFPg})~\&~(\ref{eqn:width_P_approx}) clearly overestimate the full light-front prediction (\ref{eqn:LFWF_TFFPg}), as we will see later. For charmonia, the effect is substantial. 
A further dilemma is that, for pseudoscalar quarkonia $\eta_c$, the decay constant $f_{\eta_c}$ cannot be unambiguously extracted from the experimental measurement. The BLFQ prediction $f_{\eta_c} = 0.42(7)$ GeV \cite{Li:2017mlw} is in good agreement with model-dependent value $f_{\eta_c} = 0.335(75)$ GeV extracted from the experiment \cite{CLEO:2000moj} as well as with predictions from Lattice QCD and DSE/BSE calculations \cite{Davies:2010ip, Donald:2012ga, McNeile:2012qf, Blank:2011ha}.  

The two-photon TFF of $\eta_c$ is measured by BaBar collaboration \cite{BaBar:2010siw}. The BaBar data can be well described by a monopole fit with a pole mass $\Lambda^2 = 8.5\pm 0.6 \pm 0.7 \,\mathrm{GeV}^2 \approx M_{J/\psi}^2$. This corroborates the vector meson dominant (VMD) model with the nearest vector mesons.  

The BaBar data along with the monopole fit are shown in Fig.~\ref{fig:TFF_etac_1S}. The BLFQ prediction (\ref{eqn:LFWF_TFFPg}) is in excellent agreement with the BaBar data. Following the analysis in Ref.~\cite{Li:2017mlw} for the decay constants, our calculation uses the $N_{\max}=8$ results, which corresponds to a UV resolution $\mu_\textsc{uv} \approx \kappa\sqrt{N_{\max}} = 2.8\,\mathrm{GeV}$. The basis sensitivity shown as uncertainty band is estimated as the difference between the $N_{\max}=8$ and $N_{\max}=16$ results.  

A DSE/BSE result is included for comparison \cite{Chen:2016bpj}. Two recent Lattice calculations are less successful due to the difficulty to represent the photon with large virtuality \cite{CLQCD:2016ugl, CLQCD:2020njc}. More recent Lattice calculations instead focus on the on-shell amplitude.

Note that we present the TFF $F_{P\gamma}(Q^2)$ instead of the normalized TFF 
$F_{P\gamma}(Q^2)/F_{P\gamma}(0)$ to avoid the propagation of errors from $F_{P\gamma}(0)$ present in both theory and experiment. The value $F_{P\gamma}(0) \propto \surd\Gamma_{P\to\gamma\gamma}$ can also be compared from Fig.~\ref{fig:TFF_etac_1S}, where we add the PDG value combining measurement of the two-photon width from various processes \cite{ParticleDataGroup:2020ssz}. 
The prediction using the LCDA (``BLFQ/DA'') are also included in Fig.~\ref{fig:TFF_etac_1S} for comparison. At $Q^2=0$, the prediction clearly overshoots the BLFQ prediction as expected. 

\begin{figure}
\centering
\includegraphics[width=0.43\textwidth]{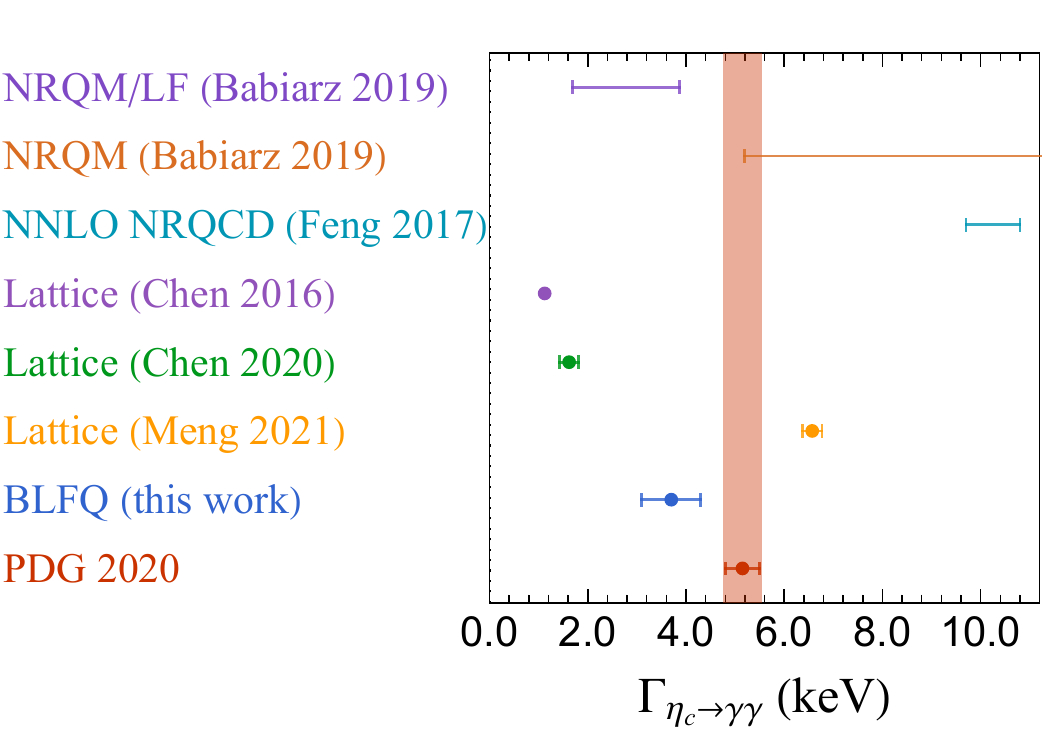} \\
\includegraphics[width=0.44\textwidth]{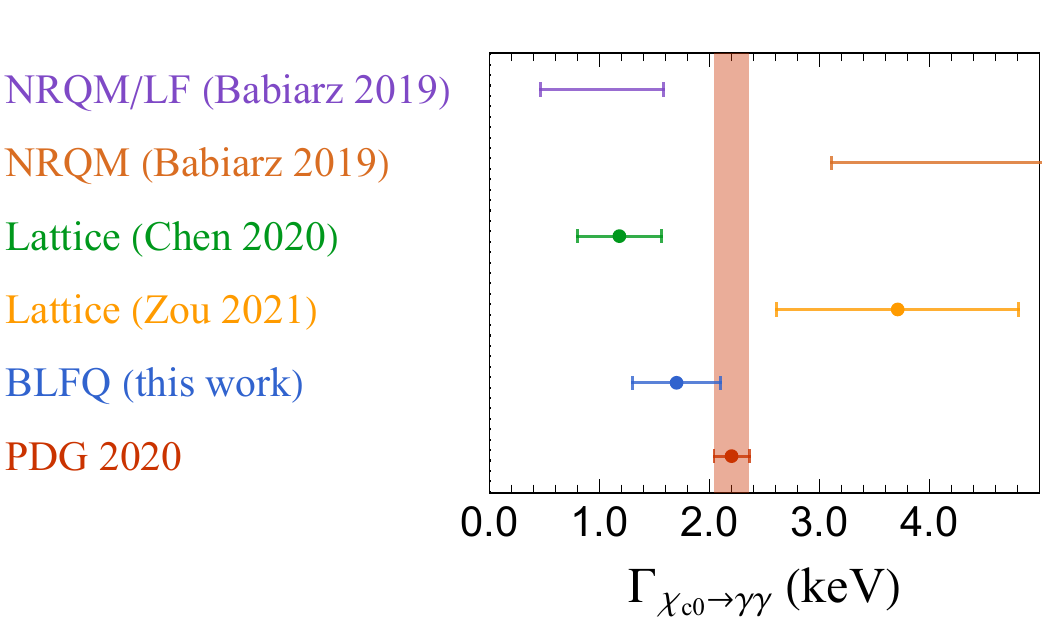}
\caption{Comparison of selected recent theoretical predictions of the two-photon transition widths for $\eta_c(1S)$ and $\chi_{c0}(1P)$ with quantified uncertainties. See Table~\ref{tab:TFF0} for more comparisons of the widths.}
\label{fig:two_photon_width}
\end{figure}

The on-shell two-photon width from PDG as well as from selected theoretical predictions are collected in Table~\ref{tab:TFF0}. A comparison of selected recent predictions with quantified uncertainties is shown in Fig.~\ref{fig:two_photon_width}. Without adjusting any parameter, our predictions appear very competitive with other theoretical approaches.

There is no measurement of the $\eta_c(2S)$ TFF at the present. Thus, our results Fig.~\ref{fig:TFF_etac_2S} are predictions. $F_{\eta_c'\gamma}(0)$ can be accessed via other processes and are also shown in Fig.~\ref{fig:TFF_etac_2S}. Our prediction is again in good agreement with the PDG value \cite{ParticleDataGroup:2020ssz}. The monopole fit with a pole mass $\Lambda=M_{\psi(2S)}$ is depicted for comparison.

In the literature, one of the major sources of uncertainty is the quark mass $m_f$. In various models,  it is known that both the shape of the TFF $F_{P\gamma}(Q^2)/F_{P\gamma}(0)$ and the two-photon width $\Gamma_{P\to\gamma\gamma}$ are sensitive to the value of the quark mass $m_f$, in opposing directions.  
When confronted by the experimental data, the former, dictated by the pole mass $\Lambda^2 \approx \langle \frac{k^2_\perp+m_f^2}{x(1-x)} \rangle$, typically favors a lower value $m_f \sim 1.3\,\mathrm{GeV}$ close to the current quark mass 
whereas the latter favors a larger value close to the effective quark mass. Indeed, (\ref{eqn:width_P_approx}) is more accurate if $m_f^2 \to m_f^2 + \langle k_\perp^2\rangle$.
In our calculation, the effective quark mass $m_f = 1.57\,\mathrm{GeV}$ is determined from the mass spectroscopy, leaving no room for parameter manipulation. 
Therefore, it is remarkable that our LFWFs can reproduce the two-photon TFF and the two-photon width simultaneously. 
 
\begin{table*}
\centering 
\caption{A compilation of experimental measurements and selected theoretical predictions of the two-photon width $\Gamma_{H\to\gamma\gamma}$ for charmonium. For $\chi_{c1}$, the reduced two-photon width $\widetilde\Gamma_{H\to\gamma\gamma}$ is listed instead. {The uncertainties have been combined in quadrature.} 
See the text for more details. }\label{tab:TFF0}
\begin{tabular}{cllllllllll}
\toprule

 & & $\eta_c(1S)$  & $\eta_c(2S)$ & $\chi_{c0}(1P)$ & $\chi_{c0}(2P)$& $\chi_{c1}(1P)$ & $\chi_{c1}(2P)$& $\chi_{c2}(1P)$ & $\chi_{c2}(2P)$  
 \\
 \colrule
\multirow{10}{*}{$\Gamma_{H\to\gamma\gamma}$} & Experiment \cite{ParticleDataGroup:2020ssz} & 5.15(35) &  2.1(1.6) & 2.20(16) & -- & -- & \uwave{0.02--0.5}\footnotemark[1] & 0.56(5) & --  \\
\multirow{11}{*}{or $\uwave{\widetilde\Gamma_{H\to\gamma\gamma}}$} & \textbf{BLFQ} &  \textbf{3.7(6)} & \textbf{1.9(4)} &  \textbf{1.7(4)} & \textbf{0.68(22)} & \textbf{\uwave{3.0(5)}} & \textbf{\uwave{3(1)}} & \textbf{0.70(13)} & \textbf{0.58(25)} 
\\
\multirow{11}{*}{($\mathrm{keV}$)} & Lattice \cite{Meng:2021ecs,Zou:2021mgf} & 6.57(20) &  -- & 3.7(1.1) & -- & -- & -- & -- & --  \\
 & Lattice \cite{CLQCD:2016ugl} & 1.122(14) &  -- & -- & -- & -- & -- & --  & -- \\
 & Lattice \cite{CLQCD:2020njc} & 1.62(19) & -- & 1.18(38) &  -- & -- & -- & -- & -- \\
 & Lattice \cite{Dudek:2006ut} & 2.65(99) & -- & 2.41(1.04) &  -- & -- & -- & -- & -- \\
 & NRQCD \cite{Feng:2017hlu} & 9.7--10.8 &  -- & -- & -- & -- & -- & -- & -- 
 \\
 & DSE/BSE \cite{Chen:2016bpj} & 6.39 & -- & 2.39 & -- & -- & -- & 0.655 & -- 
 \\
 &  LFQM \cite{Ryu:2018egt} & 4.88 &  -- & -- & -- & -- & -- &  -- & -- 
 \\
 &  LFQM \cite{Hwang:2006cua,Hwang:2010iq} & 5.7--9.7 & -- & 2.36(35) &  -- & -- & -- & 0.35(1) & -- 
 \\
  & NRQM/LF \cite{Babiarz:2019sfa,Babiarz:2020jkh}& 1.7--3.9 & 0.94--2.45 & 1.43--2.09  & -- & -- & -- & -- & --   \\
 &  NRQM \cite{Babiarz:2019sfa,Babiarz:2020jkh} & 5.2--21 & 3.1--8.8 & 3.1--5.5  & -- & -- & -- & -- & -- \\
\botrule
\end{tabular}
\\
\raggedright\footnotemark[1]{The value is for $\chi_{c1}(3872)$ \cite{Belle:2020ndp}.}
\end{table*}

\subsection{$\chi_{c0} \to \gamma\gamma$}

The transition amplitude of this process can be parametrized by two TFFs \cite{Hoferichter:2020lap, Babiarz:2020jkh, DeWitt:2003rs},
\begin{multline}\label{eq:Mnumu_scalar}
\mathcal M^{\mu\nu} =
\frac{4\pi \alpha_\mathrm{em}}{M_S^2} \Big\{M_S^2
\big[ (q_1\cdot q_2) g^{\mu\nu} - q_2^\mu q_1^\nu \big] F_{1}^S(q_1^2, q_2^2)  +\\
 \big[ q_1^2q_2^2g^{\mu\nu} + (q_1\cdot q_2) q_1^\mu q_2^\nu - q_1^2q_2^\mu q_2^\nu - q_2^2 q_1^\mu q_1^\nu \big] F_2^S(q_1^2, q_2^2) 
\Big\}.
\end{multline}
With this definition, the two-photon width is, 
\begin{equation}\label{eq:scalar_width}
\Gamma_{S\to\gamma\gamma} = \frac{\pi\alpha_\text{em}^2}{4}M_S^3 \big|F_{1}^S(0,0)\big|^2.
\end{equation}
If one photon is off-shell, as in ``single tagged'' experiments, the two-photon width $\Gamma_{\chi_{c0}\to\gamma^*\gamma}$ can be solely described by TFF $F_1^{S}$. 
Thus, it is convenient to introduce a single-variable TFF $F_{S\gamma}(q^2) = F_1^{S}(-q^2,0) =  F_1^{S}(0,-q^2)$. Its relation with the single-tag 
two-photon width $\Gamma_{\chi_{c0}\to\gamma^*\gamma}$ is,
\begin{equation}
 \Gamma_{S\to\gamma^*\gamma} = \frac{\pi \alpha^2_\text{em}}{2}\frac{(M_S^2+Q^2)^3}{M_S^3} \big|F_{S\gamma}(Q^2)\big|^2.
\end{equation}
The LFWF representation of the TFF is,

\begin{multline}\label{eqn:TFF_LFWF_S}
F_{S\gamma}(Q^2) =  e_f^22\sqrt{2N_C} \int_0^1\frac{\dd x}{2\sqrt{x(1-x)}} \int \frac{\dd^2k_\perp}{(2\pi)^3} \\
\times\Big\{
 \psi_{\uparrow\downarrow+\downarrow\uparrow/S}(x, \vec k_\perp) 
\frac{(1-2x)[x(1-x)Q^2+m^2_f]}{[k^2_\perp+x(1-x)Q^2+m_f^2]^2}
\\
+ \psi_{\uparrow\uparrow/S}(x, \vec k_\perp) \frac{\sqrt{2}m_f (k_x+ik_y)}{[k^2_\perp+x(1-x)Q^2+m_f^2]^2}
\Big\}. 
\end{multline}
Similar to the pseudoscalar case, the approximate result using the LCDA representation reads,
\begin{equation}
F_{S\gamma}(Q^2)= e_f^2 f_S \int_0^1\dd x \frac{(1-2x)\phi_S(x, \mu)}{x(1-x)Q^2+m_f^2},
\end{equation}
where the scalar meson LCDA is defined as,
\begin{equation}\label{eqn:TFF_DA_S}
 f_S\phi_S(x, \mu) = 
\sqrt{\frac{2N_C}{x(1-x)}} \int\limits^{\mu^2}\frac{\dd^2k_\perp}{(2\pi)^3} \psi_{\uparrow\downarrow+\downarrow\uparrow/S}(x, \vec k_\perp),
\end{equation}
and is normalized by $\int_0^1\dd x\, (1-2x)\phi_S(x, \mu) = 1$. Here, we only kept the leading-twist contribution.

\begin{figure}
\centering
\includegraphics[width=0.48\textwidth]{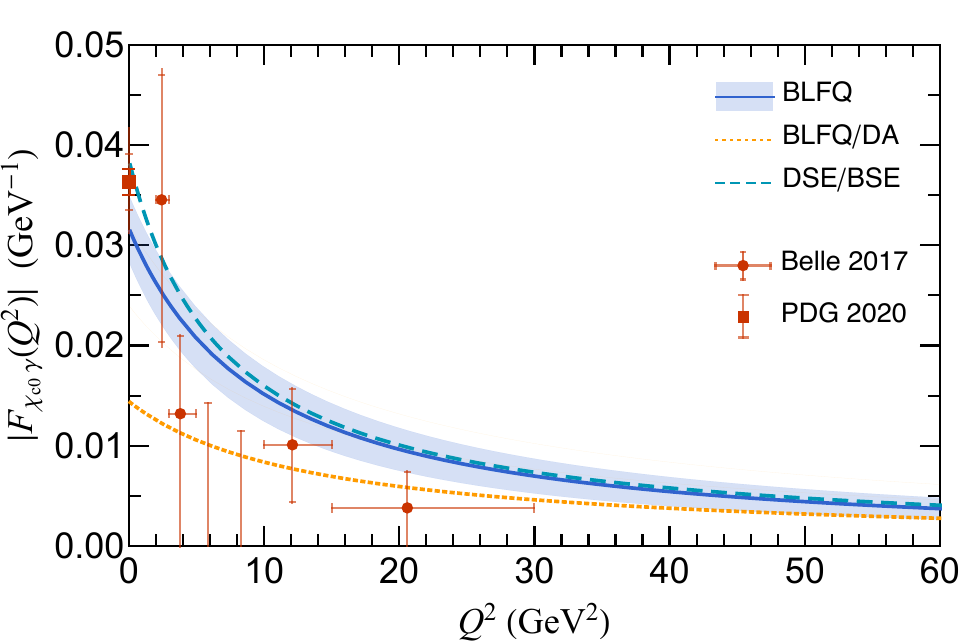}
\caption{The TFF for $\chi_{c0}(1P)$. The DSE/BSE results are included for comparison \cite{Chen:2016bpj}. }
\label{fig:TFF_chic0}
\end{figure}

The Belle collaboration provided the first measurement of the TFF $F_{\chi_{c0}\gamma}(Q^2)$, albeit with limited statistics \cite{Belle:2017xsz}. The extracted data are shown in Fig.~\ref{fig:TFF_chic0}. A recent result from DSE/BSE is also shown for comparison \cite{Chen:2016bpj}. 
The TFF at $Q^2=0$ is accessed from the two-photon decay width $\Gamma_{S\to\gamma\gamma}$ from various processes as compiled by PDG \cite{ParticleDataGroup:2020ssz}. Our predictions using the LFWFs (\ref{eqn:TFF_LFWF_S}) as well as using the LCDA (\ref{eqn:TFF_DA_S}) are both in agreement with the experimental data despite the low statistics. The BLFQ/DA prediction is scaled by  $F_{S\gamma}(0)$ obtained from BLFQ. The value is also in agreement with that extracted from the two-photon decay width $\Gamma_{S\to\gamma\gamma}$ (see Table~\ref{tab:TFF0}).

\subsection{$\chi_{c1}\to\gamma\gamma$}

The two-photon width of an axial vector $1^{++}$ vanishes due to the Landau-Yang theorem \cite{Landau:1948kw, Yang:1950rg}. Instead, one can define the reduced width {as},
\begin{equation}
\widetilde\Gamma_{A\to\gamma\gamma} = \lim_{q_1^2\to 0}\frac{M_A^2}{q_1^2} \Gamma(A \to \gamma_\textsc{l}^*\gamma_\textsc{t}). 
\end{equation}
The Belle collaboration recently measured the reduced width of $\chi_{c1}(3872)$ using single-tag events \cite{Belle:2020ndp}. The obtained result is,
$\widetilde\Gamma_{\chi_{c1}(3872)\to\gamma\gamma} = 20 - 500 \,\mathrm{eV}$.

On the theory side, the amplitude of $A \to \gamma \gamma$ can be parametrized by three TFFs, $F_{1\text{--}3}^{A}$:
\begin{equation}\label{eq:Mmunu_1pp}
\begin{split}
\mathcal {M}^{\mu\nu\alpha} = &\frac{i4\pi \alpha_\mathrm{em}}{M_A^2} \Big\{
 \epsilon^{\mu\nu\beta\gamma} q_{1\beta} q_{2\gamma} (q_1-q_2)^\alpha F_{1}^{A}(q_1^2, q_2^2) \\
+\big[ & \epsilon^{\alpha\nu\beta\gamma}q_{1\beta}q_{2\gamma}q_1^\mu + \epsilon^{\alpha\mu\nu\beta}q_{2\beta}q_1^2\big]F_{2}^{A}(q_1^2, q_2^2) \\
+\big[ & \epsilon^{\alpha\mu\beta\gamma}q_{1\beta}q_{2\gamma}q_2^\nu + \epsilon^{\alpha\mu\nu\beta}q_{1\beta}q_2^2\big]F_{3}^{A}(q_1^2, q_2^2)
\Big\},
\end{split}
\end{equation}
where $F_{2}^{A}(q_1^2,q_2^2) = -F_{3}^{A}(q_2^2,q_1^2)$ owing to the boson statistics of photons. 

The reduced width is related to the TFFs $F_{2}^{A}$ and $F_{3}^{A}$ {as},
\begin{equation}
\widetilde\Gamma_{A\to\gamma\gamma} = \frac{\pi \alpha_\text{em}^2}{6} M_A^3\big|F_{A\gamma}(0)\big|^2,
\end{equation}
where $F_{A\gamma}(q^2) = F_{2}^{A}(-q^2,0)/M_A=-F_{3}^{A}(0,-q^2)/M_A$. 

The LFWF representation of this TFF is,
\begin{multline}\label{eqn:TFF_LFWF_A}
F_{A\gamma}(Q^2) = \frac{8M_A}{M_A^2+Q^2} e_f^2\sqrt{N_C}  \int_0^1\frac{\dd x}{2\sqrt{x(1-x)}} \\ \times \int \frac{\dd^2k_\perp}{(2\pi)^3} 
\frac{(k_x+ik_y)\psi_{\uparrow\downarrow+\downarrow\uparrow/A}^{(\lambda=-1)}(x, \vec k_\perp)}{k^2_\perp+x(1-x)Q^2+m_f^2}.
\end{multline}
Using the BLFQ LFWFs, the reduced two-photon width of $\chi_{c1}(1P)$ is predicted to be $\tilde\Gamma_{\chi_{c1}\to\gamma\gamma} = (3.0\pm0.5)\,\mathrm{keV}$.
Our BLFQ calculation further predicts an excited pure $c\bar c$ axial vector meson with the mass $M_{\chi'_{c1}} = 3.948 (31) (17)\,\mathrm{GeV}$. The two-photon width of this state is predicted to be $\tilde\Gamma_{\chi'_{c1}\to\gamma\gamma} = (3\pm1)\,\mathrm{keV}$, a value significantly above the recent Belle measurement for the 2P candidate $\chi_{c1}(3872)$, suggesting that the state possesses  a large portion of non-$c\bar c$ component \cite{Belle:2020ndp}. 

\subsection{$\chi_{c2} \to \gamma\gamma$}
 
The two-photon decay width of the tensor $\chi_{c2}$ is measured by various experiments as compiled by PDG, and the average value is,
$\Gamma_{\chi_{c2}\to\gamma\gamma} = 0.56(5)\,\mathrm{keV}$ \cite{ParticleDataGroup:2020ssz}. Physically, this process is determined by two helicity amplitudes,
\begin{equation}
\Gamma_{T \to \gamma\gamma} = 
\frac{1}{16\pi}
\frac{1}{5M_T}\Big( \big|H_{++;0}\big|^2 + \big|H_{+-;+2}\big|^2 \Big).
\end{equation}
The LFWFs representation of the helicity amplitudes are,
\begin{align}
& H_{++;0} = e^2 e_f^2 \sqrt{2N_C} \int\frac{\dd x}{2[x(1-x)]^{\half[3]}}\int\frac{\dd^2k_\perp}{(2\pi)^3} \nonumber \\
& \times \Big\{
\frac{k_\perp^2(2x-1)\psi_{\uparrow\downarrow+\downarrow\uparrow/T}^{(\lambda=0)}(x, \vec k_\perp)}{k_\perp^2+x(1-x)Q^2+m_f^2} \nonumber \\
& +\frac{\sqrt{2}m_f(k_x+ik_y)\psi_{\uparrow\uparrow/T}^{(\lambda=0)}(x, \vec k_\perp)}{k_\perp^2+x(1-x)Q^2+m_f^2}
\Big\}, \label{eq:LFWF_T2pp_Hpp0} \\
& H_{+-;+2} = e^2 e_f^2 \sqrt{2N_C} \int\frac{\dd x}{2[{x(1-x)}]^{\half[3]}}\int\frac{\dd^2k_\perp}{(2\pi)^3} \nonumber \\
& \times  \Big\{
(k_x-ik_y)^2\frac{(2x-1)\psi_{\uparrow\downarrow+\downarrow\uparrow/T}^{(\lambda=+2)}(x, \vec k_\perp)+\psi_{\uparrow\downarrow-\downarrow\uparrow/T}^{(\lambda=+2)}(x, \vec k_\perp)}{k_\perp^2+x(1-x)Q^2+m_f^2} \nonumber \\
& + \frac{\sqrt{2}m_f(k_x-ik_y)\psi_{\uparrow\uparrow/T}^{(\lambda=+2)}(x, \vec k_\perp)}{k_\perp^2+x(1-x)Q^2+m_f^2} \Big\}. \label{eq:LFWF_T2pp_Hpm0}
\end{align}

From these expressions, we obtain the width {$\Gamma_{\chi_{c2}\to\gamma\gamma} = 0.70(13)\,\mathrm{keV}$}, consistent with the PDG value 0.56(5) keV.
Similarly, we can make {a} prediction for the 2P tensor as pure $c\bar c$ state. {The value $\Gamma_{\chi'_{c2}\to\gamma\gamma} = 0.58(25)\,\mathrm{keV}$ is consistent with the PDG lower bound $\Gamma_{\chi_{c2}(3930)} > 0.17 \, \text{keV}$ for the 2P candidate $\chi_{c2}(3930)$.}

\begin{figure}
\centering
\includegraphics[width=0.46\textwidth]{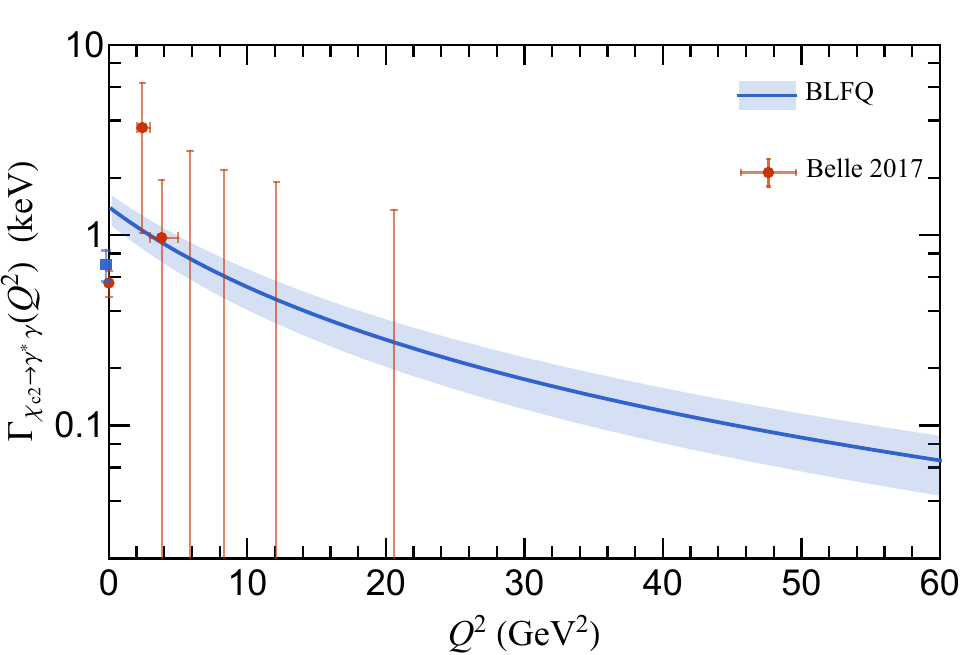}
\caption{The single tagged two-photon decay width of $\chi_{c2}(1P)$.}
\label{fig:TPW_chic2}
\end{figure}

The Belle collaboration provided the first measurement of the single tagged width $\Gamma_{\chi_{c2}\to\gamma^*\gamma}(Q^2)$, albeit with limited statistics \cite{Belle:2017xsz}. The data are compared with our result in Fig.~\ref{fig:TPW_chic2}, where our BLFQ prediction is in good agreement with the experimental measurements.

\subsection{Asymptotic limit}

\begin{figure}
    \centering
    \includegraphics[width=0.48\textwidth]{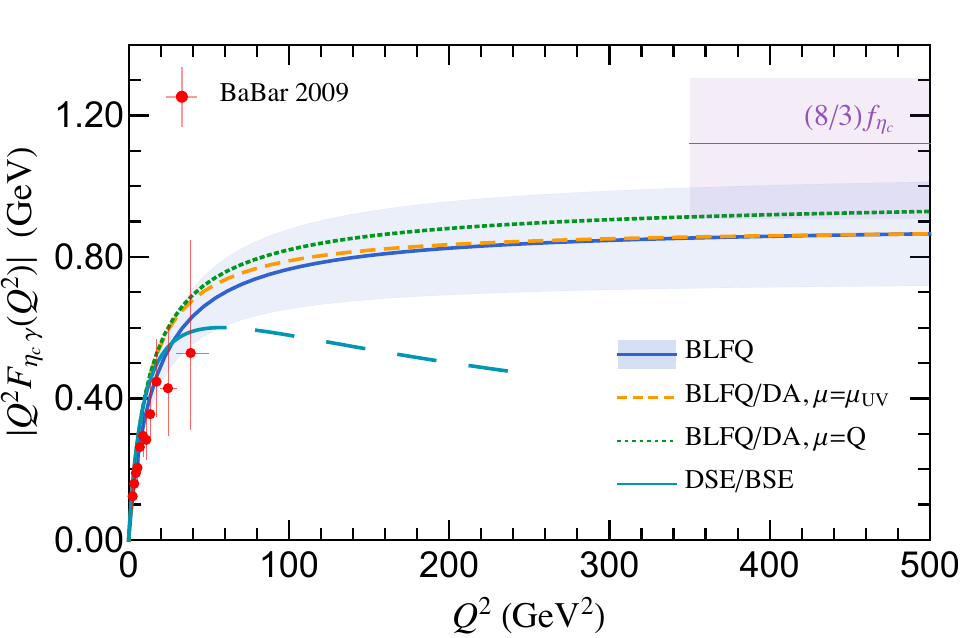}
    \caption{Large $Q^2$ behavior of the TFF. }
    \label{fig:Q2TFF}
\end{figure}

The large $Q^2$ asymptotic behavior of the TFFs can be computed from perturbative QCD \cite{Hoferichter:2020lap},
\begin{align}
& Q^2F_{P\gamma}(Q^2) \overset{Q^2\to\infty}{=} 6e_f^2f_P, \label{eqn:pQCD_P}\\
& Q^2F_{S\gamma}(Q^2) \overset{Q^2\to\infty}{=} 6e_f^2f_S(\mu). \label{eqn:pQCD_S}
\end{align}
Figure~\ref{fig:Q2TFF} shows $Q^2F_{H\gamma}$ as a function of $Q^2$ up to large $Q^2$ from various approaches. 
The BLFQ results (solid blue) and the BLFQ/DA results (dashed orange) are computed with fixed scale $\mu = \mu_\textsc{uv} \approx \kappa\sqrt{N_{\max}}$. For the pseudoscalar, their agreement at large $Q^2$ is excellent as expected. For large $Q^2$, the evolution of the DAs may not be negligible. We thus evolve the LCDA using the ERBL evolution \cite{Lepage:1980fj}. The evolved results (BLFQ/DA, $\mu=Q$, green dotted) show some small deviation from the fixed scale results. However, up to $Q^2=500\,\text{GeV}^2$ all results are below the pQCD asympotic limit, confirming the long-standing observation that the convergence to the pQCD asympotic limit is very slow \cite{Chernyak:1983ej}.

\section{Summary}

In this work, we investigated the two-photon transitions of heavy quarkonia in the light-front approach using wave functions directly computed from an effective Hamiltonian inspired by light-front holography and one-gluon exchange from light-front QCD. We computed the two photon decay widths and the transition form factors of pseudoscalar, scalar, axial vector and tensor mesons. The results are in good agreement with the available experimental measurements. This is significant since all results reported here are pure predictions -- no parameter fitting is performed to obtain these results. 
We also make predictions for two-photon processes {yet} to be measured.  

The success of our work and a similar success of the DSE/BSE calculation imply the relativistic nature of the charmonium system. For describing observables sensitive to the short-distance physics at $\sim\alpha_s M_{c\bar c}$, the relativistic formulation is required. The light-front formalism is intrinsically relativistic and has the further advantage that it is directly related to the partonic picture at large momentum transfer. 

In combination with the successes in charmonium spectroscopy as well as other observables, e.g., decay constant, radiative transitions, leptonic transitions, electromagnetic form factors, and parton distributions, our work provides a unified framework to describe the relativistic structure of the charmonium system. We also compare the results with those computed from the extracted light-cone distribution amplitudes, laying the foundation for applications {to} exclusive processes at high energy.

\section*{Acknowledgements}
The authors acknowledge valuable discussions with T. Lappi, G. Chen, X. Zhao and W. Yan. The authors would also like to thank M. Masuda for providing the original data on the two-photon transition form factor of $\chi_{c0}$ and $\chi_{c2}$. This work is supported by the New faculty start-up fund of the University of Science and Technology of China. Y.~Li and J.~P.~Vary are supported in part by the US Department of Energy (DOE) under Grant No. DE-FG02-87ER40371. M.~Li is supported by the Academy of Finland, project 321840 and  under the European Union’s Horizon 2020 research and innovation program by the European Research Council (ERC, grant agreement No. ERC-2015-CoG-681707) and by the STRONG-2020 project (grant agreement No 824093). The content of this article does not reflect the official opinion of the European Union and responsibility for the information and views expressed therein lies entirely with the authors. 
M.~Li acknowledges financial support from Xunta de Galicia (Centro singular de investigación de Galicia accreditation 2019-2022), European Union ERDF, the “María de Maeztu” Units of Excellence program, the Spanish Research State Agency, and European Research Council project ERC-2018-ADG-835105 YoctoLHC.

\bibliography{TFF}

\begin{thebibliography}{47}%
\makeatletter
\providecommand \@ifxundefined [1]{%
 \@ifx{#1\undefined}
}%
\providecommand \@ifnum [1]{%
 \ifnum #1\expandafter \@firstoftwo
 \else \expandafter \@secondoftwo
 \fi
}%
\providecommand \@ifx [1]{%
 \ifx #1\expandafter \@firstoftwo
 \else \expandafter \@secondoftwo
 \fi
}%
\providecommand \natexlab [1]{#1}%
\providecommand \enquote  [1]{``#1''}%
\providecommand \bibnamefont  [1]{#1}%
\providecommand \bibfnamefont [1]{#1}%
\providecommand \citenamefont [1]{#1}%
\providecommand \href@noop [0]{\@secondoftwo}%
\providecommand \href [0]{\begingroup \@sanitize@url \@href}%
\providecommand \@href[1]{\@@startlink{#1}\@@href}%
\providecommand \@@href[1]{\endgroup#1\@@endlink}%
\providecommand \@sanitize@url [0]{\catcode `\\12\catcode `\$12\catcode
  `\&12\catcode `\#12\catcode `\^12\catcode `\_12\catcode `\%12\relax}%
\providecommand \@@startlink[1]{}%
\providecommand \@@endlink[0]{}%
\providecommand \url  [0]{\begingroup\@sanitize@url \@url }%
\providecommand \@url [1]{\endgroup\@href {#1}{\urlprefix }}%
\providecommand \urlprefix  [0]{URL }%
\providecommand \Eprint [0]{\href }%
\providecommand \doibase [0]{https://doi.org/}%
\providecommand \selectlanguage [0]{\@gobble}%
\providecommand \bibinfo  [0]{\@secondoftwo}%
\providecommand \bibfield  [0]{\@secondoftwo}%
\providecommand \translation [1]{[#1]}%
\providecommand \BibitemOpen [0]{}%
\providecommand \bibitemStop [0]{}%
\providecommand \bibitemNoStop [0]{.\EOS\space}%
\providecommand \EOS [0]{\spacefactor3000\relax}%
\providecommand \BibitemShut  [1]{\csname bibitem#1\endcsname}%
\let\auto@bib@innerbib\@empty
\bibitem [{\citenamefont {Brambilla}\ \emph {et~al.}(2011)\citenamefont
  {Brambilla} \emph {et~al.}}]{Brambilla:2010cs}%
  \BibitemOpen
  \bibfield  {author} {\bibinfo {author} {\bibfnamefont {N.}~\bibnamefont
  {Brambilla}} \emph {et~al.},\ }\bibfield  {title} {\bibinfo {title} {{Heavy
  Quarkonium: Progress, Puzzles, and Opportunities}},\ }\href
  {https://doi.org/10.1140/epjc/s10052-010-1534-9} {\bibfield  {journal}
  {\bibinfo  {journal} {Eur. Phys. J. C}\ }\textbf {\bibinfo {volume} {71}},\
  \bibinfo {pages} {1534} (\bibinfo {year} {2011})},\ \Eprint
  {https://arxiv.org/abs/1010.5827} {arXiv:1010.5827 [hep-ph]} \BibitemShut
  {NoStop}%
\bibitem [{\citenamefont {Berger}\ and\ \citenamefont
  {Wagner}(1987)}]{Berger:1986ii}%
  \BibitemOpen
  \bibfield  {author} {\bibinfo {author} {\bibfnamefont {C.}~\bibnamefont
  {Berger}}\ and\ \bibinfo {author} {\bibfnamefont {W.}~\bibnamefont
  {Wagner}},\ }\bibfield  {title} {\bibinfo {title} {{Photon-Photon
  Reactions}},\ }\href {https://doi.org/10.1016/0370-1573(87)90012-3}
  {\bibfield  {journal} {\bibinfo  {journal} {Phys. Rept.}\ }\textbf {\bibinfo
  {volume} {146}},\ \bibinfo {pages} {1} (\bibinfo {year} {1987})}\BibitemShut
  {NoStop}%
\bibitem [{\citenamefont {Poppe}(1986)}]{Poppe:1986dq}%
  \BibitemOpen
  \bibfield  {author} {\bibinfo {author} {\bibfnamefont {M.}~\bibnamefont
  {Poppe}},\ }\bibfield  {title} {\bibinfo {title} {{Exclusive Hadron
  Production in Two Photon Reactions}},\ }\href
  {https://doi.org/10.1142/S0217751X8600023X} {\bibfield  {journal} {\bibinfo
  {journal} {Int. J. Mod. Phys. A}\ }\textbf {\bibinfo {volume} {1}},\ \bibinfo
  {pages} {545} (\bibinfo {year} {1986})}\BibitemShut {NoStop}%
\bibitem [{\citenamefont {Feindt}(1991)}]{Feindt:1991rb}%
  \BibitemOpen
  \bibfield  {author} {\bibinfo {author} {\bibfnamefont {M.}~\bibnamefont
  {Feindt}},\ }\bibfield  {title} {\bibinfo {title} {{A Review of two photon
  physics}},\ }in\ \href@noop {} {\emph {\bibinfo {booktitle} {{26th Rencontres
  de Moriond: High-energy Hadronic Interactions}}}}\ (\bibinfo {year} {1991})\
  pp.\ \bibinfo {pages} {409--421}\BibitemShut {NoStop}%
\bibitem [{\citenamefont {Lansberg}(2020)}]{Lansberg:2019adr}%
  \BibitemOpen
  \bibfield  {author} {\bibinfo {author} {\bibfnamefont {J.-P.}\ \bibnamefont
  {Lansberg}},\ }\bibfield  {title} {\bibinfo {title} {{New Observables in
  Inclusive Production of Quarkonia}},\ }\href
  {https://doi.org/10.1016/j.physrep.2020.08.007} {\bibfield  {journal}
  {\bibinfo  {journal} {Phys. Rept.}\ }\textbf {\bibinfo {volume} {889}},\
  \bibinfo {pages} {1} (\bibinfo {year} {2020})},\ \Eprint
  {https://arxiv.org/abs/1903.09185} {arXiv:1903.09185 [hep-ph]} \BibitemShut
  {NoStop}%
\bibitem [{\citenamefont {Feng}\ \emph {et~al.}(2015)\citenamefont {Feng},
  \citenamefont {Jia},\ and\ \citenamefont {Sang}}]{Feng:2015uha}%
  \BibitemOpen
  \bibfield  {author} {\bibinfo {author} {\bibfnamefont {F.}~\bibnamefont
  {Feng}}, \bibinfo {author} {\bibfnamefont {Y.}~\bibnamefont {Jia}},\ and\
  \bibinfo {author} {\bibfnamefont {W.-L.}\ \bibnamefont {Sang}},\ }\bibfield
  {title} {\bibinfo {title} {{Can Nonrelativistic QCD Explain the
  $\gamma\gamma^* \to \eta_c$ Transition Form Factor Data?}},\ }\href
  {https://doi.org/10.1103/PhysRevLett.115.222001} {\bibfield  {journal}
  {\bibinfo  {journal} {Phys. Rev. Lett.}\ }\textbf {\bibinfo {volume} {115}},\
  \bibinfo {pages} {222001} (\bibinfo {year} {2015})},\ \Eprint
  {https://arxiv.org/abs/1505.02665} {arXiv:1505.02665 [hep-ph]} \BibitemShut
  {NoStop}%
\bibitem [{\citenamefont {Feng}\ \emph {et~al.}(2017)\citenamefont {Feng},
  \citenamefont {Jia},\ and\ \citenamefont {Sang}}]{Feng:2017hlu}%
  \BibitemOpen
  \bibfield  {author} {\bibinfo {author} {\bibfnamefont {F.}~\bibnamefont
  {Feng}}, \bibinfo {author} {\bibfnamefont {Y.}~\bibnamefont {Jia}},\ and\
  \bibinfo {author} {\bibfnamefont {W.-L.}\ \bibnamefont {Sang}},\ }\bibfield
  {title} {\bibinfo {title} {{Next-to-Next-to-Leading-Order QCD Corrections to
  the Hadronic width of Pseudoscalar Quarkonium}},\ }\href
  {https://doi.org/10.1103/PhysRevLett.119.252001} {\bibfield  {journal}
  {\bibinfo  {journal} {Phys. Rev. Lett.}\ }\textbf {\bibinfo {volume} {119}},\
  \bibinfo {pages} {252001} (\bibinfo {year} {2017})},\ \Eprint
  {https://arxiv.org/abs/1707.05758} {arXiv:1707.05758 [hep-ph]} \BibitemShut
  {NoStop}%
\bibitem [{\citenamefont {Babiarz}\ \emph {et~al.}(2019)\citenamefont
  {Babiarz}, \citenamefont {Goncalves}, \citenamefont {Pasechnik},
  \citenamefont {Sch\"afer},\ and\ \citenamefont {Szczurek}}]{Babiarz:2019sfa}%
  \BibitemOpen
  \bibfield  {author} {\bibinfo {author} {\bibfnamefont {I.}~\bibnamefont
  {Babiarz}}, \bibinfo {author} {\bibfnamefont {V.~P.}\ \bibnamefont
  {Goncalves}}, \bibinfo {author} {\bibfnamefont {R.}~\bibnamefont
  {Pasechnik}}, \bibinfo {author} {\bibfnamefont {W.}~\bibnamefont
  {Sch\"afer}},\ and\ \bibinfo {author} {\bibfnamefont {A.}~\bibnamefont
  {Szczurek}},\ }\bibfield  {title} {\bibinfo {title} {{${\gamma^* \gamma^* \to
  \eta_c (1S,2S)}$ transition form factors for spacelike photons}},\ }\href
  {https://doi.org/10.1103/PhysRevD.100.054018} {\bibfield  {journal} {\bibinfo
   {journal} {Phys. Rev. D}\ }\textbf {\bibinfo {volume} {100}},\ \bibinfo
  {pages} {054018} (\bibinfo {year} {2019})},\ \Eprint
  {https://arxiv.org/abs/1908.07802} {arXiv:1908.07802 [hep-ph]} \BibitemShut
  {NoStop}%
\bibitem [{\citenamefont {Babiarz}\ \emph {et~al.}(2020)\citenamefont
  {Babiarz}, \citenamefont {Pasechnik}, \citenamefont {Sch\"afer},\ and\
  \citenamefont {Szczurek}}]{Babiarz:2020jkh}%
  \BibitemOpen
  \bibfield  {author} {\bibinfo {author} {\bibfnamefont {I.}~\bibnamefont
  {Babiarz}}, \bibinfo {author} {\bibfnamefont {R.}~\bibnamefont {Pasechnik}},
  \bibinfo {author} {\bibfnamefont {W.}~\bibnamefont {Sch\"afer}},\ and\
  \bibinfo {author} {\bibfnamefont {A.}~\bibnamefont {Szczurek}},\ }\bibfield
  {title} {\bibinfo {title} {{Hadroproduction of scalar $P$-wave quarkonia in
  the light-front k$_{T}$ -factorization approach}},\ }\href
  {https://doi.org/10.1007/JHEP06(2020)101} {\bibfield  {journal} {\bibinfo
  {journal} {JHEP}\ }\textbf {\bibinfo {volume} {06}},\ \bibinfo {pages}
  {101}},\ \Eprint {https://arxiv.org/abs/2002.09352} {arXiv:2002.09352
  [hep-ph]} \BibitemShut {NoStop}%
\bibitem [{\citenamefont {Liu}\ \emph {et~al.}(2020)\citenamefont {Liu},
  \citenamefont {Meng},\ and\ \citenamefont {Zhang}}]{Liu:2020qfz}%
  \BibitemOpen
  \bibfield  {author} {\bibinfo {author} {\bibfnamefont {C.}~\bibnamefont
  {Liu}}, \bibinfo {author} {\bibfnamefont {Y.}~\bibnamefont {Meng}},\ and\
  \bibinfo {author} {\bibfnamefont {K.-L.}\ \bibnamefont {Zhang}},\ }\bibfield
  {title} {\bibinfo {title} {{Ward identity of the vector current and the decay
  rate of $\eta_c\rightarrow\gamma\gamma$ in lattice QCD}},\ }\href
  {https://doi.org/10.1103/PhysRevD.102.034502} {\bibfield  {journal} {\bibinfo
   {journal} {Phys. Rev. D}\ }\textbf {\bibinfo {volume} {102}},\ \bibinfo
  {pages} {034502} (\bibinfo {year} {2020})},\ \Eprint
  {https://arxiv.org/abs/2004.03907} {arXiv:2004.03907 [hep-lat]} \BibitemShut
  {NoStop}%
\bibitem [{\citenamefont {Dudek}\ and\ \citenamefont
  {Edwards}(2006)}]{Dudek:2006ut}%
  \BibitemOpen
  \bibfield  {author} {\bibinfo {author} {\bibfnamefont {J.~J.}\ \bibnamefont
  {Dudek}}\ and\ \bibinfo {author} {\bibfnamefont {R.~G.}\ \bibnamefont
  {Edwards}},\ }\bibfield  {title} {\bibinfo {title} {{Two Photon Decays of
  Charmonia from Lattice QCD}},\ }\href
  {https://doi.org/10.1103/PhysRevLett.97.172001} {\bibfield  {journal}
  {\bibinfo  {journal} {Phys. Rev. Lett.}\ }\textbf {\bibinfo {volume} {97}},\
  \bibinfo {pages} {172001} (\bibinfo {year} {2006})},\ \Eprint
  {https://arxiv.org/abs/hep-ph/0607140} {arXiv:hep-ph/0607140} \BibitemShut
  {NoStop}%
\bibitem [{\citenamefont {Chen}\ \emph {et~al.}(2016)\citenamefont {Chen} \emph
  {et~al.}}]{CLQCD:2016ugl}%
  \BibitemOpen
  \bibfield  {author} {\bibinfo {author} {\bibfnamefont {T.}~\bibnamefont
  {Chen}} \emph {et~al.} (\bibinfo {collaboration} {CLQCD}),\ }\bibfield
  {title} {\bibinfo {title} {{Two-photon decays of $\eta _c$ from lattice
  QCD}},\ }\href {https://doi.org/10.1140/epjc/s10052-016-4212-8} {\bibfield
  {journal} {\bibinfo  {journal} {Eur. Phys. J. C}\ }\textbf {\bibinfo {volume}
  {76}},\ \bibinfo {pages} {358} (\bibinfo {year} {2016})},\ \Eprint
  {https://arxiv.org/abs/1602.00076} {arXiv:1602.00076 [hep-lat]} \BibitemShut
  {NoStop}%
\bibitem [{\citenamefont {Chen}\ \emph {et~al.}(2020)\citenamefont {Chen},
  \citenamefont {Gong}, \citenamefont {Li}, \citenamefont {Liu}, \citenamefont
  {Liu}, \citenamefont {Liu}, \citenamefont {Ma}, \citenamefont {Meng},
  \citenamefont {Xiong},\ and\ \citenamefont {Zhang}}]{CLQCD:2020njc}%
  \BibitemOpen
  \bibfield  {author} {\bibinfo {author} {\bibfnamefont {Y.}~\bibnamefont
  {Chen}}, \bibinfo {author} {\bibfnamefont {M.}~\bibnamefont {Gong}}, \bibinfo
  {author} {\bibfnamefont {N.}~\bibnamefont {Li}}, \bibinfo {author}
  {\bibfnamefont {C.}~\bibnamefont {Liu}}, \bibinfo {author} {\bibfnamefont
  {Y.-B.}\ \bibnamefont {Liu}}, \bibinfo {author} {\bibfnamefont
  {Z.}~\bibnamefont {Liu}}, \bibinfo {author} {\bibfnamefont {J.-P.}\
  \bibnamefont {Ma}}, \bibinfo {author} {\bibfnamefont {Y.}~\bibnamefont
  {Meng}}, \bibinfo {author} {\bibfnamefont {C.}~\bibnamefont {Xiong}},\ and\
  \bibinfo {author} {\bibfnamefont {K.-L.}\ \bibnamefont {Zhang}} (\bibinfo
  {collaboration} {CLQCD}),\ }\bibfield  {title} {\bibinfo {title} {{Lattice
  study of two-photon decay widths for scalar and pseudo-scalar charmonium}},\
  }\href {https://doi.org/10.1088/1674-1137/44/8/083108} {\bibfield  {journal}
  {\bibinfo  {journal} {Chin. Phys. C}\ }\textbf {\bibinfo {volume} {44}},\
  \bibinfo {pages} {083108} (\bibinfo {year} {2020})},\ \Eprint
  {https://arxiv.org/abs/2003.09817} {arXiv:2003.09817 [hep-lat]} \BibitemShut
  {NoStop}%
\bibitem [{\citenamefont {Meng}\ \emph {et~al.}(2021)\citenamefont {Meng},
  \citenamefont {Feng}, \citenamefont {Liu}, \citenamefont {Wang},\ and\
  \citenamefont {Zou}}]{Meng:2021ecs}%
  \BibitemOpen
  \bibfield  {author} {\bibinfo {author} {\bibfnamefont {Y.}~\bibnamefont
  {Meng}}, \bibinfo {author} {\bibfnamefont {X.}~\bibnamefont {Feng}}, \bibinfo
  {author} {\bibfnamefont {C.}~\bibnamefont {Liu}}, \bibinfo {author}
  {\bibfnamefont {T.}~\bibnamefont {Wang}},\ and\ \bibinfo {author}
  {\bibfnamefont {Z.}~\bibnamefont {Zou}},\ }\bibfield  {title} {\bibinfo
  {title} {{First-principle calculation of $\eta_c\rightarrow 2\gamma$ decay
  width from lattice QCD}},\ }\Eprint {https://arxiv.org/abs/2109.09381}
  {arXiv:2109.09381 [hep-lat]}  (\bibinfo {year} {2021})\BibitemShut {NoStop}%
\bibitem [{\citenamefont {Zou}\ \emph {et~al.}(2021)\citenamefont {Zou},
  \citenamefont {Meng},\ and\ \citenamefont {Liu}}]{Zou:2021mgf}%
  \BibitemOpen
  \bibfield  {author} {\bibinfo {author} {\bibfnamefont {Z.}~\bibnamefont
  {Zou}}, \bibinfo {author} {\bibfnamefont {Y.}~\bibnamefont {Meng}},\ and\
  \bibinfo {author} {\bibfnamefont {C.}~\bibnamefont {Liu}},\ }\bibfield
  {title} {\bibinfo {title} {{Lattice calculation of $\chi_{c0} \rightarrow
  2\gamma$ decay width}},\ }\Eprint {https://arxiv.org/abs/2111.00768}
  {arXiv:2111.00768 [hep-lat]}  (\bibinfo {year} {2021})\BibitemShut {NoStop}%
\bibitem [{\citenamefont {Chen}\ \emph
  {et~al.}(2017{\natexlab{a}})\citenamefont {Chen}, \citenamefont {Ding},
  \citenamefont {Chang},\ and\ \citenamefont {Liu}}]{Chen:2016bpj}%
  \BibitemOpen
  \bibfield  {author} {\bibinfo {author} {\bibfnamefont {J.}~\bibnamefont
  {Chen}}, \bibinfo {author} {\bibfnamefont {M.}~\bibnamefont {Ding}}, \bibinfo
  {author} {\bibfnamefont {L.}~\bibnamefont {Chang}},\ and\ \bibinfo {author}
  {\bibfnamefont {Y.-x.}\ \bibnamefont {Liu}},\ }\bibfield  {title} {\bibinfo
  {title} {{Two Photon Transition Form Factor of $\bar{c}c $ Quarkonia}},\
  }\href {https://doi.org/10.1103/PhysRevD.95.016010} {\bibfield  {journal}
  {\bibinfo  {journal} {Phys. Rev. D}\ }\textbf {\bibinfo {volume} {95}},\
  \bibinfo {pages} {016010} (\bibinfo {year} {2017}{\natexlab{a}})},\ \Eprint
  {https://arxiv.org/abs/1611.05960} {arXiv:1611.05960 [nucl-th]} \BibitemShut
  {NoStop}%
\bibitem [{\citenamefont {Brodsky}\ \emph {et~al.}(1998)\citenamefont
  {Brodsky}, \citenamefont {Pauli},\ and\ \citenamefont
  {Pinsky}}]{Brodsky:1997de}%
  \BibitemOpen
  \bibfield  {author} {\bibinfo {author} {\bibfnamefont {S.~J.}\ \bibnamefont
  {Brodsky}}, \bibinfo {author} {\bibfnamefont {H.-C.}\ \bibnamefont {Pauli}},\
  and\ \bibinfo {author} {\bibfnamefont {S.~S.}\ \bibnamefont {Pinsky}},\
  }\bibfield  {title} {\bibinfo {title} {{Quantum chromodynamics and other
  field theories on the light cone}},\ }\href
  {https://doi.org/10.1016/S0370-1573(97)00089-6} {\bibfield  {journal}
  {\bibinfo  {journal} {Phys. Rept.}\ }\textbf {\bibinfo {volume} {301}},\
  \bibinfo {pages} {299} (\bibinfo {year} {1998})},\ \Eprint
  {https://arxiv.org/abs/hep-ph/9705477} {arXiv:hep-ph/9705477} \BibitemShut
  {NoStop}%
\bibitem [{\citenamefont {Lepage}\ and\ \citenamefont
  {Brodsky}(1980)}]{Lepage:1980fj}%
  \BibitemOpen
  \bibfield  {author} {\bibinfo {author} {\bibfnamefont {G.~P.}\ \bibnamefont
  {Lepage}}\ and\ \bibinfo {author} {\bibfnamefont {S.~J.}\ \bibnamefont
  {Brodsky}},\ }\bibfield  {title} {\bibinfo {title} {{Exclusive Processes in
  Perturbative Quantum Chromodynamics}},\ }\href
  {https://doi.org/10.1103/PhysRevD.22.2157} {\bibfield  {journal} {\bibinfo
  {journal} {Phys. Rev. D}\ }\textbf {\bibinfo {volume} {22}},\ \bibinfo
  {pages} {2157} (\bibinfo {year} {1980})}\BibitemShut {NoStop}%
\bibitem [{\citenamefont {Chernyak}\ and\ \citenamefont
  {Zhitnitsky}(1984)}]{Chernyak:1983ej}%
  \BibitemOpen
  \bibfield  {author} {\bibinfo {author} {\bibfnamefont {V.~L.}\ \bibnamefont
  {Chernyak}}\ and\ \bibinfo {author} {\bibfnamefont {A.~R.}\ \bibnamefont
  {Zhitnitsky}},\ }\bibfield  {title} {\bibinfo {title} {{Asymptotic Behavior
  of Exclusive Processes in QCD}},\ }\href
  {https://doi.org/10.1016/0370-1573(84)90126-1} {\bibfield  {journal}
  {\bibinfo  {journal} {Phys. Rept.}\ }\textbf {\bibinfo {volume} {112}},\
  \bibinfo {pages} {173} (\bibinfo {year} {1984})}\BibitemShut {NoStop}%
\bibitem [{\citenamefont {Beuf}(2016)}]{Beuf:2016wdz}%
  \BibitemOpen
  \bibfield  {author} {\bibinfo {author} {\bibfnamefont {G.}~\bibnamefont
  {Beuf}},\ }\bibfield  {title} {\bibinfo {title} {{Dipole factorization for
  DIS at NLO: Loop correction to the $\gamma^*_{T,L}\to q\overline q$
  light-front wave functions}},\ }\href
  {https://doi.org/10.1103/PhysRevD.94.054016} {\bibfield  {journal} {\bibinfo
  {journal} {Phys. Rev. D}\ }\textbf {\bibinfo {volume} {94}},\ \bibinfo
  {pages} {054016} (\bibinfo {year} {2016})},\ \Eprint
  {https://arxiv.org/abs/1606.00777} {arXiv:1606.00777 [hep-ph]} \BibitemShut
  {NoStop}%
\bibitem [{\citenamefont {Chernyak}\ and\ \citenamefont
  {Eidelman}(2015)}]{Chernyak:2014wra}%
  \BibitemOpen
  \bibfield  {author} {\bibinfo {author} {\bibfnamefont {V.~L.}\ \bibnamefont
  {Chernyak}}\ and\ \bibinfo {author} {\bibfnamefont {S.~I.}\ \bibnamefont
  {Eidelman}},\ }\bibfield  {title} {\bibinfo {title} {{Hard exclusive two
  photon processes in QCD}},\ }\href
  {https://doi.org/10.1016/j.ppnp.2014.09.002} {\bibfield  {journal} {\bibinfo
  {journal} {Prog. Part. Nucl. Phys.}\ }\textbf {\bibinfo {volume} {80}},\
  \bibinfo {pages} {1} (\bibinfo {year} {2015})},\ \Eprint
  {https://arxiv.org/abs/1409.3348} {arXiv:1409.3348 [hep-ph]} \BibitemShut
  {NoStop}%
\bibitem [{\citenamefont {Li}\ \emph {et~al.}(2017)\citenamefont {Li},
  \citenamefont {Maris},\ and\ \citenamefont {Vary}}]{Li:2017mlw}%
  \BibitemOpen
  \bibfield  {author} {\bibinfo {author} {\bibfnamefont {Y.}~\bibnamefont
  {Li}}, \bibinfo {author} {\bibfnamefont {P.}~\bibnamefont {Maris}},\ and\
  \bibinfo {author} {\bibfnamefont {J.~P.}\ \bibnamefont {Vary}},\ }\bibfield
  {title} {\bibinfo {title} {{Quarkonium as a relativistic bound state on the
  light front}},\ }\href {https://doi.org/10.1103/PhysRevD.96.016022}
  {\bibfield  {journal} {\bibinfo  {journal} {Phys. Rev. D}\ }\textbf {\bibinfo
  {volume} {96}},\ \bibinfo {pages} {016022} (\bibinfo {year} {2017})},\
  \Eprint {https://arxiv.org/abs/1704.06968} {arXiv:1704.06968 [hep-ph]}
  \BibitemShut {NoStop}%
\bibitem [{\citenamefont {Li}(2017)}]{Li:2017data}%
  \BibitemOpen
  \bibfield  {author} {\bibinfo {author} {\bibfnamefont {Y.}~\bibnamefont
  {Li}},\ }\bibfield  {title} {\bibinfo {title} {{Quarkonium light-front wave
  functions}},\ }\bibfield  {journal} {\bibinfo  {journal} {Mendeley Data}\
  }\textbf {\bibinfo {volume} {V1}},\ \href
  {https://doi.org/10.17632/5bgp37xwz4.1} {10.17632/5bgp37xwz4.1} (\bibinfo
  {year} {2017})\BibitemShut {NoStop}%
\bibitem [{\citenamefont {Lees}\ \emph {et~al.}(2010)\citenamefont {Lees} \emph
  {et~al.}}]{BaBar:2010siw}%
  \BibitemOpen
  \bibfield  {author} {\bibinfo {author} {\bibfnamefont {J.~P.}\ \bibnamefont
  {Lees}} \emph {et~al.} (\bibinfo {collaboration} {BaBar}),\ }\bibfield
  {title} {\bibinfo {title} {{Measurement of the $\gamma \gamma* --> \eta_c$
  transition form factor}},\ }\href
  {https://doi.org/10.1103/PhysRevD.81.052010} {\bibfield  {journal} {\bibinfo
  {journal} {Phys. Rev. D}\ }\textbf {\bibinfo {volume} {81}},\ \bibinfo
  {pages} {052010} (\bibinfo {year} {2010})},\ \Eprint
  {https://arxiv.org/abs/1002.3000} {arXiv:1002.3000 [hep-ex]} \BibitemShut
  {NoStop}%
\bibitem [{\citenamefont {Masuda}\ \emph {et~al.}(2018)\citenamefont {Masuda}
  \emph {et~al.}}]{Belle:2017xsz}%
  \BibitemOpen
  \bibfield  {author} {\bibinfo {author} {\bibfnamefont {M.}~\bibnamefont
  {Masuda}} \emph {et~al.} (\bibinfo {collaboration} {Belle}),\ }\bibfield
  {title} {\bibinfo {title} {{Study of $K^0_S$ pair production in single-tag
  two-photon collisions}},\ }\href {https://doi.org/10.1103/PhysRevD.97.052003}
  {\bibfield  {journal} {\bibinfo  {journal} {Phys. Rev. D}\ }\textbf {\bibinfo
  {volume} {97}},\ \bibinfo {pages} {052003} (\bibinfo {year} {2018})},\
  \Eprint {https://arxiv.org/abs/1712.02145} {arXiv:1712.02145 [hep-ex]}
  \BibitemShut {NoStop}%
\bibitem [{\citenamefont {Teramoto}\ \emph {et~al.}(2021)\citenamefont
  {Teramoto} \emph {et~al.}}]{Belle:2020ndp}%
  \BibitemOpen
  \bibfield  {author} {\bibinfo {author} {\bibfnamefont {Y.}~\bibnamefont
  {Teramoto}} \emph {et~al.} (\bibinfo {collaboration} {Belle}),\ }\bibfield
  {title} {\bibinfo {title} {{Evidence for $X(3872)\rightarrow J/\psi
  \pi^+\pi^-$ Produced in Single-Tag Two-Photon Interactions}},\ }\href
  {https://doi.org/10.1103/PhysRevLett.126.122001} {\bibfield  {journal}
  {\bibinfo  {journal} {Phys. Rev. Lett.}\ }\textbf {\bibinfo {volume} {126}},\
  \bibinfo {pages} {122001} (\bibinfo {year} {2021})},\ \Eprint
  {https://arxiv.org/abs/2007.05696} {arXiv:2007.05696 [hep-ex]} \BibitemShut
  {NoStop}%
\bibitem [{\citenamefont {Zyla}\ \emph {et~al.}(2020)\citenamefont {Zyla} \emph
  {et~al.}}]{ParticleDataGroup:2020ssz}%
  \BibitemOpen
  \bibfield  {author} {\bibinfo {author} {\bibfnamefont {P.~A.}\ \bibnamefont
  {Zyla}} \emph {et~al.} (\bibinfo {collaboration} {Particle Data Group}),\
  }\bibfield  {title} {\bibinfo {title} {{Review of Particle Physics}},\ }\href
  {https://doi.org/10.1093/ptep/ptaa104} {\bibfield  {journal} {\bibinfo
  {journal} {PTEP}\ }\textbf {\bibinfo {volume} {2020}},\ \bibinfo {pages}
  {083C01} (\bibinfo {year} {2020})}\BibitemShut {NoStop}%
\bibitem [{\citenamefont {Li}\ \emph {et~al.}(2018)\citenamefont {Li},
  \citenamefont {Li}, \citenamefont {Maris},\ and\ \citenamefont
  {Vary}}]{Li:2018uif}%
  \BibitemOpen
  \bibfield  {author} {\bibinfo {author} {\bibfnamefont {M.}~\bibnamefont
  {Li}}, \bibinfo {author} {\bibfnamefont {Y.}~\bibnamefont {Li}}, \bibinfo
  {author} {\bibfnamefont {P.}~\bibnamefont {Maris}},\ and\ \bibinfo {author}
  {\bibfnamefont {J.~P.}\ \bibnamefont {Vary}},\ }\bibfield  {title} {\bibinfo
  {title} {{Radiative transitions between $0^{-+}$ and $1^{--}$ heavy quarkonia
  on the light front}},\ }\href {https://doi.org/10.1103/PhysRevD.98.034024}
  {\bibfield  {journal} {\bibinfo  {journal} {Phys. Rev. D}\ }\textbf {\bibinfo
  {volume} {98}},\ \bibinfo {pages} {034024} (\bibinfo {year} {2018})},\
  \Eprint {https://arxiv.org/abs/1803.11519} {arXiv:1803.11519 [hep-ph]}
  \BibitemShut {NoStop}%
\bibitem [{\citenamefont {Tang}\ \emph {et~al.}(2021)\citenamefont {Tang},
  \citenamefont {Jia}, \citenamefont {Maris},\ and\ \citenamefont
  {Vary}}]{Tang:2020org}%
  \BibitemOpen
  \bibfield  {author} {\bibinfo {author} {\bibfnamefont {S.}~\bibnamefont
  {Tang}}, \bibinfo {author} {\bibfnamefont {S.}~\bibnamefont {Jia}}, \bibinfo
  {author} {\bibfnamefont {P.}~\bibnamefont {Maris}},\ and\ \bibinfo {author}
  {\bibfnamefont {J.~P.}\ \bibnamefont {Vary}},\ }\bibfield  {title} {\bibinfo
  {title} {{Semileptonic decay of Bc to \ensuremath{\eta}c and
  J/\ensuremath{\psi} on the light front}},\ }\href
  {https://doi.org/10.1103/PhysRevD.104.016002} {\bibfield  {journal} {\bibinfo
   {journal} {Phys. Rev. D}\ }\textbf {\bibinfo {volume} {104}},\ \bibinfo
  {pages} {016002} (\bibinfo {year} {2021})},\ \Eprint
  {https://arxiv.org/abs/2011.05454} {arXiv:2011.05454 [hep-ph]} \BibitemShut
  {NoStop}%
\bibitem [{\citenamefont {Adhikari}\ \emph {et~al.}(2019)\citenamefont
  {Adhikari}, \citenamefont {Li}, \citenamefont {Li},\ and\ \citenamefont
  {Vary}}]{Adhikari:2018umb}%
  \BibitemOpen
  \bibfield  {author} {\bibinfo {author} {\bibfnamefont {L.}~\bibnamefont
  {Adhikari}}, \bibinfo {author} {\bibfnamefont {Y.}~\bibnamefont {Li}},
  \bibinfo {author} {\bibfnamefont {M.}~\bibnamefont {Li}},\ and\ \bibinfo
  {author} {\bibfnamefont {J.~P.}\ \bibnamefont {Vary}},\ }\bibfield  {title}
  {\bibinfo {title} {{Form factors and generalized parton distributions of
  heavy quarkonia in basis light front quantization}},\ }\href
  {https://doi.org/10.1103/PhysRevC.99.035208} {\bibfield  {journal} {\bibinfo
  {journal} {Phys. Rev. C}\ }\textbf {\bibinfo {volume} {99}},\ \bibinfo
  {pages} {035208} (\bibinfo {year} {2019})},\ \Eprint
  {https://arxiv.org/abs/1809.06475} {arXiv:1809.06475 [hep-ph]} \BibitemShut
  {NoStop}%
\bibitem [{\citenamefont {Lan}\ \emph {et~al.}(2020)\citenamefont {Lan},
  \citenamefont {Mondal}, \citenamefont {Li}, \citenamefont {Li}, \citenamefont
  {Tang}, \citenamefont {Zhao},\ and\ \citenamefont {Vary}}]{Lan:2019img}%
  \BibitemOpen
  \bibfield  {author} {\bibinfo {author} {\bibfnamefont {J.}~\bibnamefont
  {Lan}}, \bibinfo {author} {\bibfnamefont {C.}~\bibnamefont {Mondal}},
  \bibinfo {author} {\bibfnamefont {M.}~\bibnamefont {Li}}, \bibinfo {author}
  {\bibfnamefont {Y.}~\bibnamefont {Li}}, \bibinfo {author} {\bibfnamefont
  {S.}~\bibnamefont {Tang}}, \bibinfo {author} {\bibfnamefont {X.}~\bibnamefont
  {Zhao}},\ and\ \bibinfo {author} {\bibfnamefont {J.~P.}\ \bibnamefont
  {Vary}},\ }\bibfield  {title} {\bibinfo {title} {{Parton Distribution
  Functions of Heavy Mesons on the Light Front}},\ }\href
  {https://doi.org/10.1103/PhysRevD.102.014020} {\bibfield  {journal} {\bibinfo
   {journal} {Phys. Rev. D}\ }\textbf {\bibinfo {volume} {102}},\ \bibinfo
  {pages} {014020} (\bibinfo {year} {2020})},\ \Eprint
  {https://arxiv.org/abs/1911.11676} {arXiv:1911.11676 [nucl-th]} \BibitemShut
  {NoStop}%
\bibitem [{\citenamefont {Chen}\ \emph
  {et~al.}(2017{\natexlab{b}})\citenamefont {Chen}, \citenamefont {Li},
  \citenamefont {Maris}, \citenamefont {Tuchin},\ and\ \citenamefont
  {Vary}}]{Chen:2016dlk}%
  \BibitemOpen
  \bibfield  {author} {\bibinfo {author} {\bibfnamefont {G.}~\bibnamefont
  {Chen}}, \bibinfo {author} {\bibfnamefont {Y.}~\bibnamefont {Li}}, \bibinfo
  {author} {\bibfnamefont {P.}~\bibnamefont {Maris}}, \bibinfo {author}
  {\bibfnamefont {K.}~\bibnamefont {Tuchin}},\ and\ \bibinfo {author}
  {\bibfnamefont {J.~P.}\ \bibnamefont {Vary}},\ }\bibfield  {title} {\bibinfo
  {title} {{Diffractive charmonium spectrum in high energy collisions in the
  basis light-front quantization approach}},\ }\href
  {https://doi.org/10.1016/j.physletb.2017.04.024} {\bibfield  {journal}
  {\bibinfo  {journal} {Phys. Lett. B}\ }\textbf {\bibinfo {volume} {769}},\
  \bibinfo {pages} {477} (\bibinfo {year} {2017}{\natexlab{b}})},\ \Eprint
  {https://arxiv.org/abs/1610.04945} {arXiv:1610.04945 [nucl-th]} \BibitemShut
  {NoStop}%
\bibitem [{\citenamefont {Chen}\ \emph {et~al.}(2019)\citenamefont {Chen},
  \citenamefont {Li}, \citenamefont {Tuchin},\ and\ \citenamefont
  {Vary}}]{Chen:2018vdw}%
  \BibitemOpen
  \bibfield  {author} {\bibinfo {author} {\bibfnamefont {G.}~\bibnamefont
  {Chen}}, \bibinfo {author} {\bibfnamefont {Y.}~\bibnamefont {Li}}, \bibinfo
  {author} {\bibfnamefont {K.}~\bibnamefont {Tuchin}},\ and\ \bibinfo {author}
  {\bibfnamefont {J.~P.}\ \bibnamefont {Vary}},\ }\bibfield  {title} {\bibinfo
  {title} {{Heavy quarkonia production at energies available at the CERN Large
  Hadron Collider and future electron-ion colliding facilities using basis
  light-front quantization wave functions}},\ }\href
  {https://doi.org/10.1103/PhysRevC.100.025208} {\bibfield  {journal} {\bibinfo
   {journal} {Phys. Rev. C}\ }\textbf {\bibinfo {volume} {100}},\ \bibinfo
  {pages} {025208} (\bibinfo {year} {2019})},\ \Eprint
  {https://arxiv.org/abs/1811.01782} {arXiv:1811.01782 [nucl-th]} \BibitemShut
  {NoStop}%
\bibitem [{\citenamefont {Landau}(1948)}]{Landau:1948kw}%
  \BibitemOpen
  \bibfield  {author} {\bibinfo {author} {\bibfnamefont {L.~D.}\ \bibnamefont
  {Landau}},\ }\bibfield  {title} {\bibinfo {title} {{On the angular momentum
  of a system of two photons}},\ }\href
  {https://doi.org/10.1016/B978-0-08-010586-4.50070-5} {\bibfield  {journal}
  {\bibinfo  {journal} {Dokl. Akad. Nauk SSSR}\ }\textbf {\bibinfo {volume}
  {60}},\ \bibinfo {pages} {207} (\bibinfo {year} {1948})}\BibitemShut
  {NoStop}%
\bibitem [{\citenamefont {Yang}(1950)}]{Yang:1950rg}%
  \BibitemOpen
  \bibfield  {author} {\bibinfo {author} {\bibfnamefont {C.-N.}\ \bibnamefont
  {Yang}},\ }\bibfield  {title} {\bibinfo {title} {{Selection Rules for the
  Dematerialization of a Particle Into Two Photons}},\ }\href
  {https://doi.org/10.1103/PhysRev.77.242} {\bibfield  {journal} {\bibinfo
  {journal} {Phys. Rev.}\ }\textbf {\bibinfo {volume} {77}},\ \bibinfo {pages}
  {242} (\bibinfo {year} {1950})}\BibitemShut {NoStop}%
\bibitem [{\citenamefont {Lappi}\ \emph {et~al.}(2020)\citenamefont {Lappi},
  \citenamefont {M\"antysaari},\ and\ \citenamefont
  {Penttala}}]{Lappi:2020ufv}%
  \BibitemOpen
  \bibfield  {author} {\bibinfo {author} {\bibfnamefont {T.}~\bibnamefont
  {Lappi}}, \bibinfo {author} {\bibfnamefont {H.}~\bibnamefont
  {M\"antysaari}},\ and\ \bibinfo {author} {\bibfnamefont {J.}~\bibnamefont
  {Penttala}},\ }\bibfield  {title} {\bibinfo {title} {{Relativistic
  corrections to the vector meson light front wave function}},\ }\href
  {https://doi.org/10.1103/PhysRevD.102.054020} {\bibfield  {journal} {\bibinfo
   {journal} {Phys. Rev. D}\ }\textbf {\bibinfo {volume} {102}},\ \bibinfo
  {pages} {054020} (\bibinfo {year} {2020})},\ \Eprint
  {https://arxiv.org/abs/2006.02830} {arXiv:2006.02830 [hep-ph]} \BibitemShut
  {NoStop}%
\bibitem [{\citenamefont {Hoferichter}\ and\ \citenamefont
  {Stoffer}(2020)}]{Hoferichter:2020lap}%
  \BibitemOpen
  \bibfield  {author} {\bibinfo {author} {\bibfnamefont {M.}~\bibnamefont
  {Hoferichter}}\ and\ \bibinfo {author} {\bibfnamefont {P.}~\bibnamefont
  {Stoffer}},\ }\bibfield  {title} {\bibinfo {title} {{Asymptotic behavior of
  meson transition form factors}},\ }\href
  {https://doi.org/10.1007/JHEP05(2020)159} {\bibfield  {journal} {\bibinfo
  {journal} {JHEP}\ }\textbf {\bibinfo {volume} {05}},\ \bibinfo {pages}
  {159}},\ \Eprint {https://arxiv.org/abs/2004.06127} {arXiv:2004.06127
  [hep-ph]} \BibitemShut {NoStop}%
\bibitem [{\citenamefont {Li}\ \emph {et~al.}(2021)\citenamefont {Li},
  \citenamefont {Li}, \citenamefont {Chen}, \citenamefont {Lappi},\ and\
  \citenamefont {Vary}}]{Li:2021cwv}%
  \BibitemOpen
  \bibfield  {author} {\bibinfo {author} {\bibfnamefont {M.}~\bibnamefont
  {Li}}, \bibinfo {author} {\bibfnamefont {Y.}~\bibnamefont {Li}}, \bibinfo
  {author} {\bibfnamefont {G.}~\bibnamefont {Chen}}, \bibinfo {author}
  {\bibfnamefont {T.}~\bibnamefont {Lappi}},\ and\ \bibinfo {author}
  {\bibfnamefont {J.~P.}\ \bibnamefont {Vary}},\ }\bibfield  {title} {\bibinfo
  {title} {{Light-front wavefunctions of mesons by design}},\ }\Eprint
  {https://arxiv.org/abs/2111.07087} {arXiv:2111.07087 [hep-ph]}  (\bibinfo
  {year} {2021})\BibitemShut {NoStop}%
\bibitem [{\citenamefont {Edwards}\ \emph {et~al.}(2001)\citenamefont {Edwards}
  \emph {et~al.}}]{CLEO:2000moj}%
  \BibitemOpen
  \bibfield  {author} {\bibinfo {author} {\bibfnamefont {K.~W.}\ \bibnamefont
  {Edwards}} \emph {et~al.} (\bibinfo {collaboration} {CLEO}),\ }\bibfield
  {title} {\bibinfo {title} {{Study of B decays to charmonium states B
  ---\ensuremath{>} eta(c) K and B ---\ensuremath{>} chi(c0) K}},\ }\href
  {https://doi.org/10.1103/PhysRevLett.86.30} {\bibfield  {journal} {\bibinfo
  {journal} {Phys. Rev. Lett.}\ }\textbf {\bibinfo {volume} {86}},\ \bibinfo
  {pages} {30} (\bibinfo {year} {2001})},\ \Eprint
  {https://arxiv.org/abs/hep-ex/0007012} {arXiv:hep-ex/0007012} \BibitemShut
  {NoStop}%
\bibitem [{\citenamefont {Davies}\ \emph {et~al.}(2010)\citenamefont {Davies},
  \citenamefont {McNeile}, \citenamefont {Follana}, \citenamefont {Lepage},
  \citenamefont {Na},\ and\ \citenamefont {Shigemitsu}}]{Davies:2010ip}%
  \BibitemOpen
  \bibfield  {author} {\bibinfo {author} {\bibfnamefont {C.~T.~H.}\
  \bibnamefont {Davies}}, \bibinfo {author} {\bibfnamefont {C.}~\bibnamefont
  {McNeile}}, \bibinfo {author} {\bibfnamefont {E.}~\bibnamefont {Follana}},
  \bibinfo {author} {\bibfnamefont {G.~P.}\ \bibnamefont {Lepage}}, \bibinfo
  {author} {\bibfnamefont {H.}~\bibnamefont {Na}},\ and\ \bibinfo {author}
  {\bibfnamefont {J.}~\bibnamefont {Shigemitsu}},\ }\bibfield  {title}
  {\bibinfo {title} {{Update: Precision $D_s$ decay constant from full lattice
  QCD using very fine lattices}},\ }\href
  {https://doi.org/10.1103/PhysRevD.82.114504} {\bibfield  {journal} {\bibinfo
  {journal} {Phys. Rev. D}\ }\textbf {\bibinfo {volume} {82}},\ \bibinfo
  {pages} {114504} (\bibinfo {year} {2010})},\ \Eprint
  {https://arxiv.org/abs/1008.4018} {arXiv:1008.4018 [hep-lat]} \BibitemShut
  {NoStop}%
\bibitem [{\citenamefont {Donald}\ \emph {et~al.}(2012)\citenamefont {Donald},
  \citenamefont {Davies}, \citenamefont {Dowdall}, \citenamefont {Follana},
  \citenamefont {Hornbostel}, \citenamefont {Koponen}, \citenamefont {Lepage},\
  and\ \citenamefont {McNeile}}]{Donald:2012ga}%
  \BibitemOpen
  \bibfield  {author} {\bibinfo {author} {\bibfnamefont {G.~C.}\ \bibnamefont
  {Donald}}, \bibinfo {author} {\bibfnamefont {C.~T.~H.}\ \bibnamefont
  {Davies}}, \bibinfo {author} {\bibfnamefont {R.~J.}\ \bibnamefont {Dowdall}},
  \bibinfo {author} {\bibfnamefont {E.}~\bibnamefont {Follana}}, \bibinfo
  {author} {\bibfnamefont {K.}~\bibnamefont {Hornbostel}}, \bibinfo {author}
  {\bibfnamefont {J.}~\bibnamefont {Koponen}}, \bibinfo {author} {\bibfnamefont
  {G.~P.}\ \bibnamefont {Lepage}},\ and\ \bibinfo {author} {\bibfnamefont
  {C.}~\bibnamefont {McNeile}},\ }\bibfield  {title} {\bibinfo {title}
  {{Precision tests of the $J/{\psi}$ from full lattice QCD: mass, leptonic
  width and radiative decay rate to ${\eta}_c$}},\ }\href
  {https://doi.org/10.1103/PhysRevD.86.094501} {\bibfield  {journal} {\bibinfo
  {journal} {Phys. Rev. D}\ }\textbf {\bibinfo {volume} {86}},\ \bibinfo
  {pages} {094501} (\bibinfo {year} {2012})},\ \Eprint
  {https://arxiv.org/abs/1208.2855} {arXiv:1208.2855 [hep-lat]} \BibitemShut
  {NoStop}%
\bibitem [{\citenamefont {McNeile}\ \emph {et~al.}(2012)\citenamefont
  {McNeile}, \citenamefont {Davies}, \citenamefont {Follana}, \citenamefont
  {Hornbostel},\ and\ \citenamefont {Lepage}}]{McNeile:2012qf}%
  \BibitemOpen
  \bibfield  {author} {\bibinfo {author} {\bibfnamefont {C.}~\bibnamefont
  {McNeile}}, \bibinfo {author} {\bibfnamefont {C.~T.~H.}\ \bibnamefont
  {Davies}}, \bibinfo {author} {\bibfnamefont {E.}~\bibnamefont {Follana}},
  \bibinfo {author} {\bibfnamefont {K.}~\bibnamefont {Hornbostel}},\ and\
  \bibinfo {author} {\bibfnamefont {G.~P.}\ \bibnamefont {Lepage}},\ }\bibfield
   {title} {\bibinfo {title} {{Heavy meson masses and decay constants from
  relativistic heavy quarks in full lattice QCD}},\ }\href
  {https://doi.org/10.1103/PhysRevD.86.074503} {\bibfield  {journal} {\bibinfo
  {journal} {Phys. Rev. D}\ }\textbf {\bibinfo {volume} {86}},\ \bibinfo
  {pages} {074503} (\bibinfo {year} {2012})},\ \Eprint
  {https://arxiv.org/abs/1207.0994} {arXiv:1207.0994 [hep-lat]} \BibitemShut
  {NoStop}%
\bibitem [{\citenamefont {Blank}\ and\ \citenamefont
  {Krassnigg}(2011)}]{Blank:2011ha}%
  \BibitemOpen
  \bibfield  {author} {\bibinfo {author} {\bibfnamefont {M.}~\bibnamefont
  {Blank}}\ and\ \bibinfo {author} {\bibfnamefont {A.}~\bibnamefont
  {Krassnigg}},\ }\bibfield  {title} {\bibinfo {title} {{Bottomonium in a
  Bethe-Salpeter-equation study}},\ }\href
  {https://doi.org/10.1103/PhysRevD.84.096014} {\bibfield  {journal} {\bibinfo
  {journal} {Phys. Rev. D}\ }\textbf {\bibinfo {volume} {84}},\ \bibinfo
  {pages} {096014} (\bibinfo {year} {2011})},\ \Eprint
  {https://arxiv.org/abs/1109.6509} {arXiv:1109.6509 [hep-ph]} \BibitemShut
  {NoStop}%
\bibitem [{\citenamefont {Ryu}\ \emph {et~al.}(2018)\citenamefont {Ryu},
  \citenamefont {Choi},\ and\ \citenamefont {Ji}}]{Ryu:2018egt}%
  \BibitemOpen
  \bibfield  {author} {\bibinfo {author} {\bibfnamefont {H.-Y.}\ \bibnamefont
  {Ryu}}, \bibinfo {author} {\bibfnamefont {H.-M.}\ \bibnamefont {Choi}},\ and\
  \bibinfo {author} {\bibfnamefont {C.-R.}\ \bibnamefont {Ji}},\ }\bibfield
  {title} {\bibinfo {title} {{Systematic twist expansion of
  $(\eta_c,\eta_b)\to\gamma^*\gamma$ transition form factors in light-front
  quark model}},\ }\href {https://doi.org/10.1103/PhysRevD.98.034018}
  {\bibfield  {journal} {\bibinfo  {journal} {Phys. Rev. D}\ }\textbf {\bibinfo
  {volume} {98}},\ \bibinfo {pages} {034018} (\bibinfo {year} {2018})},\
  \Eprint {https://arxiv.org/abs/1804.08287} {arXiv:1804.08287 [hep-ph]}
  \BibitemShut {NoStop}%
\bibitem [{\citenamefont {Hwang}\ and\ \citenamefont
  {Wei}(2007)}]{Hwang:2006cua}%
  \BibitemOpen
  \bibfield  {author} {\bibinfo {author} {\bibfnamefont {C.-W.}\ \bibnamefont
  {Hwang}}\ and\ \bibinfo {author} {\bibfnamefont {Z.-T.}\ \bibnamefont
  {Wei}},\ }\bibfield  {title} {\bibinfo {title} {{Covariant light-front
  approach for heavy quarkonium: Decay constants, P ---\ensuremath{>} gamma
  gamma and V ---\ensuremath{>} P gamma}},\ }\href
  {https://doi.org/10.1088/0954-3899/34/4/008} {\bibfield  {journal} {\bibinfo
  {journal} {J. Phys. G}\ }\textbf {\bibinfo {volume} {34}},\ \bibinfo {pages}
  {687} (\bibinfo {year} {2007})},\ \Eprint
  {https://arxiv.org/abs/hep-ph/0609036} {arXiv:hep-ph/0609036} \BibitemShut
  {NoStop}%
\bibitem [{\citenamefont {Hwang}\ and\ \citenamefont
  {Guo}(2010)}]{Hwang:2010iq}%
  \BibitemOpen
  \bibfield  {author} {\bibinfo {author} {\bibfnamefont {C.-W.}\ \bibnamefont
  {Hwang}}\ and\ \bibinfo {author} {\bibfnamefont {R.-S.}\ \bibnamefont
  {Guo}},\ }\bibfield  {title} {\bibinfo {title} {{Two-photon and two-gluon
  decays of p-wave heavy quarkonium using a covariant light-front approach}},\
  }\href {https://doi.org/10.1103/PhysRevD.82.034021} {\bibfield  {journal}
  {\bibinfo  {journal} {Phys. Rev. D}\ }\textbf {\bibinfo {volume} {82}},\
  \bibinfo {pages} {034021} (\bibinfo {year} {2010})},\ \Eprint
  {https://arxiv.org/abs/1005.2811} {arXiv:1005.2811 [hep-ph]} \BibitemShut
  {NoStop}%
\bibitem [{\citenamefont {DeWitt}\ \emph {et~al.}(2003)\citenamefont {DeWitt},
  \citenamefont {Choi},\ and\ \citenamefont {Ji}}]{DeWitt:2003rs}%
  \BibitemOpen
  \bibfield  {author} {\bibinfo {author} {\bibfnamefont {M.~A.}\ \bibnamefont
  {DeWitt}}, \bibinfo {author} {\bibfnamefont {H.-M.}\ \bibnamefont {Choi}},\
  and\ \bibinfo {author} {\bibfnamefont {C.-R.}\ \bibnamefont {Ji}},\
  }\bibfield  {title} {\bibinfo {title} {{Radiative scalar meson decays in the
  light front quark model}},\ }\href
  {https://doi.org/10.1103/PhysRevD.68.054026} {\bibfield  {journal} {\bibinfo
  {journal} {Phys. Rev. D}\ }\textbf {\bibinfo {volume} {68}},\ \bibinfo
  {pages} {054026} (\bibinfo {year} {2003})},\ \Eprint
  {https://arxiv.org/abs/hep-ph/0306060} {arXiv:hep-ph/0306060} \BibitemShut
  {NoStop}%
\end{thebibliography}%
\end{document}